\documentclass[default]{sn-jnl}

\jyear{2023}%

\theoremstyle{thmstyleone}%
\theoremstyle{thmstyletwo}%

\theoremstyle{thmstylethree}%

\makeatletter
\newcommand{\figcaption}[1]{\def\@captype{figure}\caption{#1}}
\newcommand{\tblcaption}[1]{\def\@captype{table}\caption{#1}}
\makeatother

\usepackage{caption}
\usepackage{color}
\usepackage{ulem}
\usepackage{rotating}
\usepackage{booktabs}
\usepackage{amsmath}
\usepackage{xcolor}
\usepackage{graphicx}
\usepackage{subcaption}
\usepackage{pifont}
\usepackage{multirow}

\begin{document}

\title[Automatic Hip OA Grading from CT]{Automatic hip osteoarthritis grading with uncertainty estimation from computed tomography using digitally-reconstructed radiographs}



\author*[1]{\fnm{Masachika} \sur{Masuda}}\email{masuda.masachika.mp2@is.naist.jp}

\author*[1]{\fnm{Mazen} \sur{Soufi}}\email{msoufi@is.naist.jp}
\author[1]{\fnm{Yoshito} \sur{Otake}}

\author[2]{\fnm{Keisuke} \sur{Uemura}}

\author[2]{\fnm{Sotaro} \sur{Kono}}

\author[2]{\fnm{Kazuma} \sur{Takashima}}

\author[3]{\fnm{Hidetoshi} \sur{Hamada}}

\author[1]{\fnm{Yi} \sur{Gu}}

\author[4]{\fnm{Masaki} \sur{Takao}}

\author[2]{\fnm{Seiji} \sur{Okada}} 

\author[3]{\fnm{Nobuhiko} \sur{Sugano}} 

\author*[1]{\fnm{Yoshinobu} \sur{Sato}} \email{yoshi@is.naist.jp}






\affil*[1]{\orgdiv{Division of Information Science, Graduate School of Science and Technology}, \orgname{Nara Institute of Science and Technology}, \orgaddress{\city{Ikoma}, \state{Nara}, \country{Japan}}}

\affil[2]{\orgdiv{Department of Orthopaedics, Graduate School of Medicine}, \orgname{Osaka University}, \orgaddress{\city{Suita}, \state{Osaka}, \country{Japan}}}

\affil[3]{\orgdiv{Department of Orthopaedic Medical Engineering, Graduate School of Medicine}, \orgname{Osaka University}, \orgaddress{\city{Suita}, \state{Osaka}, \country{Japan}}}

\affil[4]{\orgdiv{Department of Bone and Joint Surgery, Graduate School of Medicine}, \orgname{Ehime University}, \orgaddress{\city{Toon}, \state{Ehime}, \country{Japan}}}






\abstract{
  \textbf{Purpose:} 
    Progression of hip osteoarthritis (hip OA) leads to pain and disability, likely leading to surgical treatment such as hip arthroplasty at the terminal stage. The severity of hip OA is often classified using the Crowe and Kellgren-Lawrence (KL) classifications. However, as the classification is subjective, we aimed to develop an automated approach to classify the disease severity based on the two grades using digitally-reconstructed radiographs (DRRs) from CT images.
 
  \textbf{Methods:} 
    Automatic grading of the hip OA severity was performed using deep learning-based models. The models were trained to predict the disease grade using two grading schemes, i.e., predicting the Crowe and KL grades separately, and predicting a new ordinal label combining both grades and representing the disease progression of hip OA.  The models were trained in classification and regression settings. In addition, the model uncertainty was estimated and validated as a predictor of classification accuracy. The models were trained and validated on a database of 197 hip OA patients,  and externally validated on 52 patients. The model accuracy was evaluated using exact class accuracy (ECA), one-neighbor class accuracy (ONCA), and balanced accuracy.

  \textbf{Results:} 
      The deep learning models produced a comparable accuracy of approximately 0.65 (ECA) and 0.95 (ONCA) in the classification and regression settings.   The model uncertainty was significantly larger in cases with large classification errors (P\textless 6e-3).
 
  \textbf{Conclusion:}
  In this study, an automatic approach for grading hip OA severity from CT images was developed.  The models have shown comparable performance with high ONCA, which facilitates automated grading in large-scale CT databases and indicates the potential for further disease progression analysis.  Classification accuracy was correlated with the model uncertainty, which would allow for the prediction of classification errors. The code will be made publicly available at https://github.com/NAIST-ICB/HipOA-Grading.
}

\keywords{Hip Osteoarthritis, Crowe Grading, Kellgren and Lawrence Grading, VisionTransformer, VGG, DenseNet,  Uncertainty}



\maketitle
\section{Introduction}
\label{sec:sec_introduction}
Hip osteoarthritis (hip OA) is an increasingly prevalent disease \cite{hoy2014global}. The disease can be caused by multiple factors, including weight or trauma, or due to the acetabulum or femoral head dysplasia, such as developmental dysplasia of the hip (DDH). As OA progresses, it leads to pain and deterioration in daily life activities, making it a target for surgical treatment, including total hip arthroplasty. This necessitates a method to evaluate the progression and morphology of OA.

\begin{figure}[b]
  \centering
  \includegraphics[width=10cm]{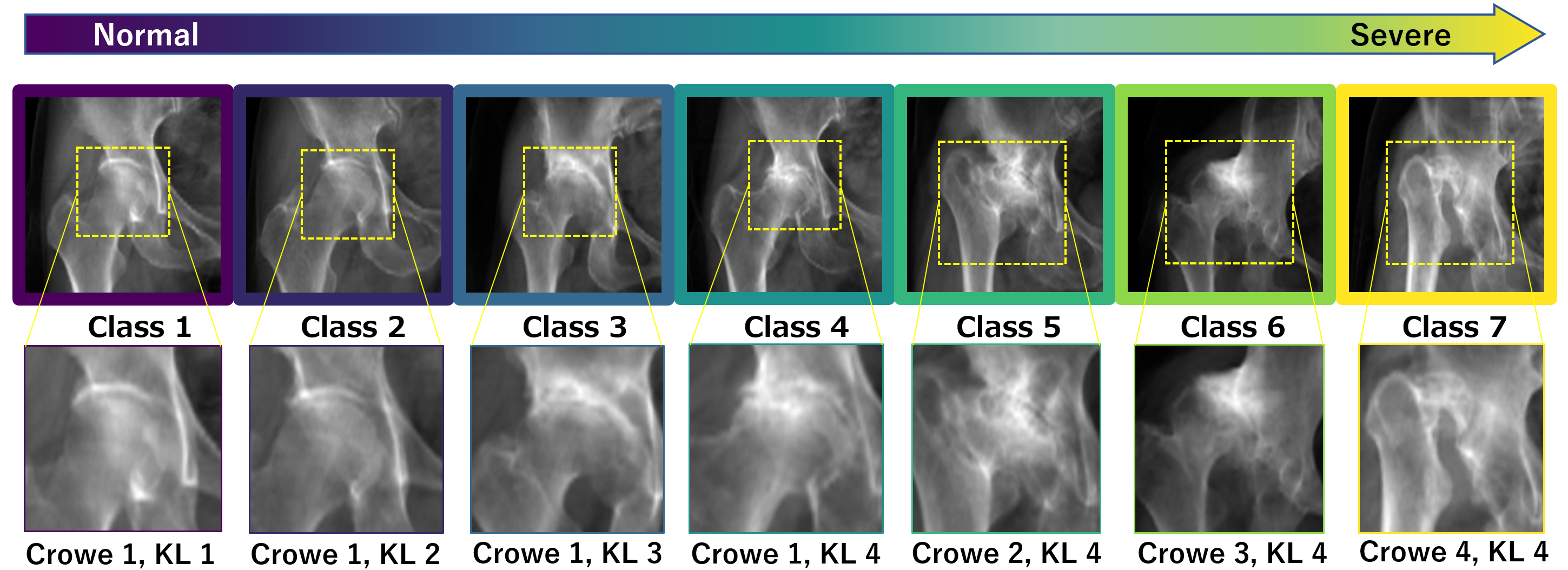}
  \caption{
    Disease grading used in the paper.
    DRR images representing the variations accompanying hip OA disease progression are depicted. The progression grades were constructed as combinations of Crowe and Kellgren and Lawrence (KL) gradings. Higher severity grades are accompanied by narrower space between the femoral head and acetabulum or sub-dislocation or dislocation of the femoral head from the acetabulum. The reason why this definition of disease classes was used will be explained in Section \ref{sec:subsec_datasets}.
  }
  \label{fig:figure01}
\end{figure}

The disease is manifested as a joint space narrowing and a deformation of the femoral head, with possible dislocation in its severe stages. Its diagnosis is usually based on X-ray radiographs and requires the expertise of orthopedic surgeons or radiologists to grade the hip deformity and disease progression. To grade hip OA, Crowe grading, i.e., the degree of femoral head dislocation from the acetabulum, and Kellgren-Lawrence (KL) grading, i.e., the degree of abrasion of the cartilage in the gap between the acetabulum and femoral head, are usually used. Figure \ref{fig:figure01} shows the different stages of disease severity with corresponding Crowe and KL stages in CT-based digitally-reconstructed radiographs (DRRs). Major challenges in the current hip OA diagnosis are subjectivity and high dependency on the surgeon, which may introduce inter- and intra-observer variability \cite{gunther1999reliability, damen2014inter}. Therefore, automated grading methods can help facilitate the diagnosis, improve reproducibility, and analyze large-scale databases. 
In particular, several studies have applied grading to X-ray images \cite{ureten2020detection, von2020development}. The reason for using CT images instead of X-ray images in our study is our interest in performing the disease progression analysis of a large-scale database of pre-operative CT images (more than 2000 cases). Since KL grading cannot be directly applied to 3D in studies using clinical CT \cite{turmezei2014new}, we extracted the 2D DRRs images for hip OA grading, as proposed in previous researches \cite{turmezei2014new, gebre2022detecting}. 

\indent Deep learning models, such as ResNet \cite{he2016deep}, VGG \cite{simonyan2014very}, and DenseNet \cite{huang2017densely}, have been successfully applied for medical image classification tasks, including KL grading of knee osteoarthritis (Knee OA) \cite{joseph2022machine, guan2022deep}, and hip OA \cite{gebre2022detecting,von2020development}. However, hip OA diagnosis was simplified into a binary classification, (normal vs. diseased) problem. Recently, transformer models have been widely used in medical image analysis tasks \cite{he2023transforrev}. A recent study has reported that the novel VisionTransformer (ViT) model \cite{dosovitskiy2020image} is capable of capturing five levels of severity changes in knee OA \cite{Konwer_2022_CVPR}. However, ViT has not been validated in hip OA grading, and the automated Crowe grading has not been considered in previous studies. 

This research aims to develop an automated grading approach that considers the disease progression stages simultaneously represented by Crowe and KL grading. Generally, deep learning models are dealt with as black boxes due to the huge number of parameters and complicated architectures. Explaining the model performance and confidence in final outputs is desirable in diagnosis. Given the possibility of misclassification by the automated approach, especially in large-scale databases with wide disease variations, a tool for assessing the model uncertainty is also investigated in this study. 

The novelty and contributions of our research are as follows; 
\begin{itemize}
  \item Development of an automated approach for grading hip OA based on Crowe and KL grading representing the disease progression rather than a binary classification model. 
  \item Investigating the potential of three deep learning models in grading hip OA in regression and classification settings.
  \item Estimating the model uncertainty and investigating its relationship with the classification accuracy.
\end{itemize}

\section{Related works}
\label{sec:sec_related_works}

Several studies have investigated the application of convolutional neural networks (CNNs) in hip OA classification. Gebre et al. trained ResNet18 on CT image-based DRRs for hip OA classification, obtaining an accuracy of 82.2\%\cite{gebre2022detecting}. Ureten et al. showed that the VGG16 trained only on X-ray images could classify hip OA with an accuracy of 90.2\% \cite{ureten2020detection}. Schacky et al. reported the classification of five hip OA features using a multi-task DenseNet \cite{von2020development}. However, in these methods, hip OA classification was treated as a binary classification, which would not allow for assessing the disease severity. Instead, our study addresses hip OA as a multi-class classification representing the disease progression stages shown in Fig. \ref{fig:figure01}.

One drawback of convolutional layers in CNNs is the inability to represent long-range dependencies in the images \cite{vaswani2017attention}. Recently, a convolution-free deep learning model, i.e., VisionTransformer (ViT) \cite{dosovitskiy2020image}, was proposed for image classification tasks. This model employs the attention mechanism \cite{vaswani2017attention}, which enables capturing global image features. Konwer et al. proposed a classification approach of knee OA into 5 severity levels using ViT \cite{Konwer_2022_CVPR}. However, to our knowledge, the potential of ViT models on hip OA classification has not been investigated yet. Furthermore, the model uncertainty has not been investigated in OA classification. Given the large-scale models and wide disease variations, estimating the model uncertainty would help understand the stability of the model against perturbations in the model weights. Model uncertainty was also correlated with the prediction accuracy in image segmentation problems \cite{hiasa2019automated}. Therefore, we investigated the potential of the model uncertainty in the hip OA classification problem and its relationship with classification accuracy.

\section{Methods}
\label{sec:sec_methods}

\subsection{Overall workflow}
\label{sec:subsec_overall_workflow}
\begin{figure}[tb]
  \centering
  \includegraphics[width=11cm]{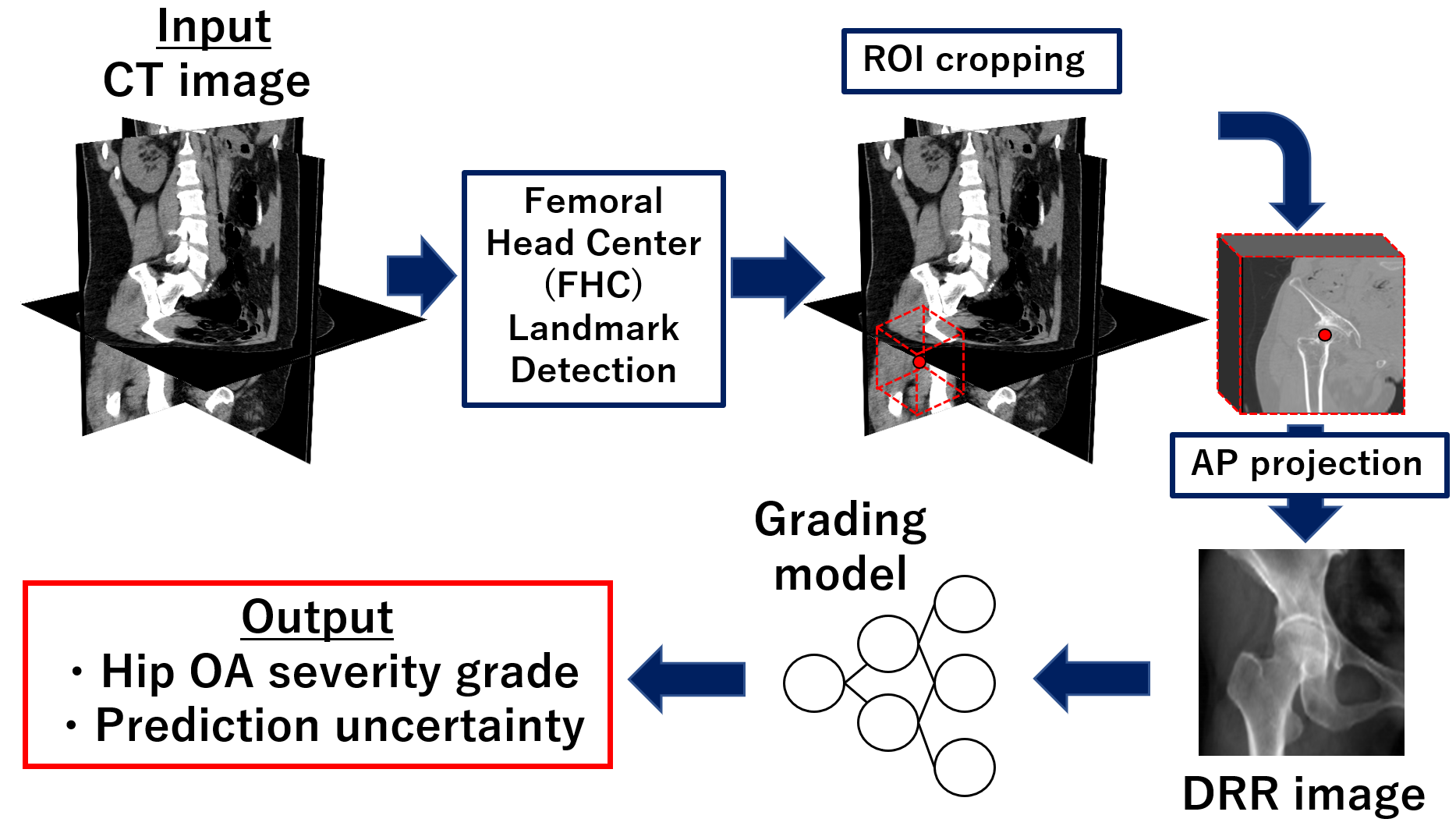}
  \caption{
    Overview of the proposed method. Hip OA severity grade was automatically predicted based on the DRR image of the hip joint region automatically 
    extracted from the CT image.
  }
  \label{fig:figure02}
\end{figure}

Figure \ref{fig:figure02} shows an overview of the proposed method for the grading of hip OA based on CT images. The femoral head centers (FHCs) were automatically detected from the CT image. FHC landmark was used to crop a region of interest (ROI), including the hip joint. A DRR image was obtained by projecting the cropped image in the anterior-posterior (AP) direction. A grading model was used to predict the hip OA severity grade based on the DRR image, and the model uncertainty was also estimated. 

\subsection{DRR image generation}
\label{sec:subsec_drr_image_generation}
In this study, unilateral DRR images of the hip joint were used. The CT images were assumed to include the pelvis-to-knee or whole lower limb regions. To limit the analysis to the hip joint region, the FHCs were detected from the CT image using a landmark detection approach.  A pre-trained landmark detection model based on 3D CNN with U-Net architecture \cite{cciccek20163d} was used to predict the right and left FHC landmarks. More details about the landmark detection can be found in \cite{uemura2023development}.  A $150 \,\rm{mm^{3}}$ cubic region centered on the FHC landmark was extracted. A projection of the extracted region in the AP direction was computed. The pixel values were normalized within the range [0, 1].

\subsection{Automated hip OA grading}
\label{sec:subsec_automated_hip_oa_grading}
In this study, three model architectures were investigated for hip OA grading. The models included CNN-based architectures, i.e., VGG \cite{simonyan2014very} and Densenet \cite{huang2017densely}, and transformer-based architecture, i.e., VisionTransformer \cite{dosovitskiy2020image}. The architectures were VGG16, DenseNet161 and VisionTransformer\_Base16 for the VGG, DenseNet and ViT models, respectively. The number of model parameters was 138M, 28M and 86M, respectively.  At the training, each model was trained in classification and regression settings.  The models were trained to predict the Crowe and KL grades in a combined or separated scheme (See Section \ref{sec:subsec_grade_labeling}). For the combined classification, the dimension of the final layer was changed from 1000 to 7. For the separated classification, two heads of fully-connected layers were added, and the dimension of the final layer was set to 4 to fit the severity levels in each grade. A softmax function was applied to the output predictions.  For the combined and separated regression, fully-connected layers were added. The dimension of the final layer was set to 1.   The output value was rounded to the closest integer and was used as the final prediction.

\subsection{Grade labeling}
\label{sec:subsec_grade_labeling}
In order to assess the impact of the labeling scheme, i.e., the prediction of both grades combined into a single label versus separated, two designs of the classification head were attempted. For the \textit{combined} prediction, models with a single classification head predicting one of the seven classes in Fig. \ref{fig:figure01} was implemented. For the \textit{separated} prediction, models with two classification heads were implemented. Particularly, Crowe and KL grades were predicted separately each with a value between 1 and 4 as

\begin{equation}
  \begin{aligned}
    {\rm Crowe(\mathit{x})} = \underset{c\in{{1,2,3,4}}}{\rm{argmax}}\mathit{(p^{Crowe}_{c}(x; \theta, \dot{\theta}))}\\
    {\rm KL(\mathit{x})} = \underset{c\in{{1,2,3,4}}}{\rm{argmax}}\mathit{(p^{KL}_{c}(x; \theta, \ddot{\theta}))}\\
  \end{aligned}
  \label{eq:2head_model_equation}
\end{equation}

\noindent where $p_c$ represents the output softmax probability of the grading head, $x$ is the input DRR image, $\theta$ denotes the parameters of the shared feature extractor, and $\dot{\theta}$, $\ddot{\theta}$ are the parameters of the Crowe and KL grading heads, respectively. The output grades by each head were determined as the class $c$ that yielded the highest probability. \\

\indent The two-head network was optimized by minimizing the loss function

\begin{equation}
  \begin{aligned}
     \mathcal{L} = \alpha\mathcal{L}_{Crowe}+\beta\mathcal{L}_{KL}\\
  \end{aligned}
  \label{eq:2head_loss_equation}
\end{equation}

\noindent where $\mathcal{L}_{Crowe}$ and $\mathcal{L}_{Crowe}$ are the losses by each head, and $\alpha$ and $\beta$ are scaling factors for Crowe and KL grading heads, respectively. The factors were adjusted based on an ablation experiment. Particularly, setting both factors to 1 led to a bias in the Crowe head towards the class $c=1$ with the large number of cases. Therefore, $\beta$ was fixed to 1, and multiple values of $\alpha$ for a larger penalty on Crowe classification loss were attempted. The values yielding the largest accuracy were selected for the 15-pattern experiment. Specifically, $\alpha=2$ was used for the classification setting of the three models, and $\alpha=7, 35$ and $35$ for the regression of the ViT, VGG, and DenseNet models, respectively.  If the model in the \textit{separated} setting outputs a combination that does not exist in Fig. \ref{fig:figure01}, the case would be considered a false prediction at the comparison with the \textit{combined} setting.

\subsection{Uncertainty estimation}
\label{sec:subsec_uncertainty_estimation}
Given the huge numbers of model parameters (See Section \ref{sec:subsec_automated_hip_oa_grading}), the stability of the models against perturbations in the model weights, i.e., epistemic uncertainty, was estimated. Monte-Carlo Dropout (MCdropout) \cite{gal2016dropout}, a simple yet efficient approach based on multiple dropout samples at inference time, was used. The MCdropout was implemented by inserting dropout layers into the grading models and activating it at inference time. The position of the dropout layer and its rate were determined experimentally. For ViT model, the default dropout layers with a dropout rate 0.1 were used \cite{dosovitskiy2020image}. In VGG, a dropout layer was inserted after the final activation (ReLU) layer at each resolution, and the rate was 0.3 for classification and 0.1 for regression. In DenseNet, dropout layers were added after each transition layer (convolution + pooling), with a rate of 0.2 for both classification and regression. The ablation experiment of the dropout rates is shown in the Appendix A Fig. \ref{fig:appendix_dropout}.\\

\indent The uncertainty was given as the variance estimated by
\begin{equation}
  \begin{split}
    {\rm Variance} ={\frac{1}{T}\sum^{T}_{i=1}({\rm Softmax}(f(x;\theta_{i}))-\Bar{y})^{2}}
  \end{split}
  \label{eq:uncertainty_equation}
\end{equation}

\noindent where $x$ is the input DRR image, $T$ is the number of dropout samples, $\Bar{y}$ is the average of the outputs obtained by dropout sampling, and $\theta_{i}$ is the parameter set corresponding to the sample $i$. In this study, $T$ was set to 50.

\subsection{Evaluation metrics}
\label{sec:subsec_evaluation_metrics}
\indent\indent\textbf{Grading accuracy}. The grading accuracy was assessed with exact class accuracy (ECA) and one-neighbor class accuracy (ONCA). ECA was computed as the ratio of the true predictions to the number of DRR images. The ONCA was computed by considering false predictions lying within one-class neighbors to the true classes as true predictions.  A model performance that is more practical and objective, taking into account class imbalances, was assessed for datasets completely independent of the training. Specifically, balanced accuracy, defined as the average of class-wise true positive rates, was also reported for both exact class and one-neighbor class predictions. 

\textbf{Regression error.} In regression, the standard error of the regression (SE) was used to evaluate the model performance. The error was calculated as 

\begin{equation}
  \begin{split}
    {\rm SE} =\frac{1}{N}\sum^{N}_{i=1}\|\hat{y_{i}}-y_{i}\|
  \end{split}
  \label{eq:standard_regression_error_equation}
\end{equation}

\noindent where $N$ is the number of DRR images, $\hat{y_{i}}$ is the predicted class, and $y_{i}$ is the ground-truth class of each image.

\subsection{Statistical analysis}
The mean and standard deviation (SD) of the evaluation metrics and uncertainty were reported. To assess the statistical significance, Student's t-test and the Mann-Whitney U-test were used for paired and unpaired measurements, respectively, with a significance level $\alpha=0.05$. The adjustment for multiple comparisons between p-values was performed using Bonferroni correction.\\

\section{Experiments}
\label{sec:sec_experiments}

\subsection{Datasets}
\label{sec:subsec_datasets}
In this study,  an internal database of  394 unilateral DRR images, generated from CT images of the hip region of 197 hip OA patients, were used for training and testing of the grading models in cross-validation experiment. The data was collected from Osaka University Hospital. The patients included 169 females and 28 males aged 61±13.5 years (mean ± standard deviation). The ratio of the primary and secondary hip OA patients was 22.3\% and 77.7\%, respectively.  An external database including 104 DRRs of 52 patients was used for testing. The images were collected from the same institution and included 40 females and 12 males aged 59.0±11.1 years.\\ 
\indent\textbf{Ground-truth hip OA grades.} Table \ref{tb:datasets} summarizes the image characteristics and disease grades. Each DRR was assigned Crowe and KL grades by an orthopedic surgeon with six years of experience.  KL 1 has ambiguous OA characteristics that are hard to distinguish from KL 0 \cite{gebre2022detecting}. Moreover, only a small number of healthy hips without OA with KL 0 were observed in our datasets. Therefore, KL 0 and 1 were merged into a single grade (KL 1).  Crowe and KL grades were combined into one class, encoding one of seven combinations. Crowe and KL grades, which are related to joint narrowing and dislocation, respectively, are independent indicators. Compared to other ethnic groups, Japanese people have significantly shallower acetabular (hip joint) depth, and higher incidence of secondary OA caused by DDH \cite{inoue2000prevalence}.  The development of hip dysplasia into dislocation has been reported \cite{hadley1990effects}. In order to investigate the capability of the deep learning model to learn the progression from joint narrowing into dislocation, an ordinal label representing the progression based on Crowe and KL grades was attempted  (See Figure \ref{fig:figure01}). Cases with no joint stenosis but high dislocation (Crowe 2, KL 1), even though possible, were not present in our database, and thus were not represented.

\begin{table}[tb]
  \centering
  \captionsetup{justification=centering}
  \caption{Details of the image characteristics and disease grades.}
  \begin{tabular}{@{}lll@{}}
    \toprule
    \multicolumn{3}{c}{Image characteristics}\\
    \midrule
    Image size of DRR (pixel) & \multicolumn{2}{c}{150 $\times$ 150}\\
    \midrule
    \multicolumn{3}{c}{Patient characteristics of the study population}\\
    \midrule
    Dataset     & Internal training/testing                   & External testing\\
    Number of cases, (hips) & 197 (394)                                               &  52 (104)\\
    Female, N(\%)           & 169 (85.8) &  40 (76.9)\\
    Male, N(\%)             & 28 (14.2)  &  12 (23.1)\\
    Mean age (SD)           & 61 (±13.5)                                              & 59 (±11.1)\\
    Primary, N(\%)          & 44 (22.3)                                               & -- \\
    Secondary, N(\%)        & 153 (77.7)                                              & --\\
    Institution             & \multicolumn{2}{c}{Osaka University Hospital}\\
    Number of classes       & \multicolumn{2}{c}{7}\\
    \midrule
    \multicolumn{3}{c}{Distribution of the disease grade}\\
    \midrule
    Class &  \multicolumn{2}{c}{Number of  hips (\%)}\\
    1 (Crowe 1, KL 1) & 112 (28) &  26 (25) \\
    2 (Crowe 1, KL 2) & 59 (15)  &  15 (14) \\
    3 (Crowe 1, KL 3) & 47 (12)  &  14 (13) \\
    4 (Crowe 1, KL 4) & 141 (36) &  22 (21) \\
    5 (Crowe 2, KL 4) & 18 (5)   &  9 (9) \\
    6 (Crowe 3, KL 4) & 12 (3)   &  8 (8) \\
    7 (Crowe 4, KL 4) & 5 (1)    &  10 (10) \\
    \bottomrule
  \end{tabular}
\label{tb:datasets}
\end{table}

%
%

\subsection{Experimental setup}
\label{sec:subsec_experimental_setup}
\indent\indent\textbf{Grading models.}  The models were trained and tested in 4-fold cross-validation experiments.  In each fold,  DRRs were randomly separated patient-wise into training, validation, and testing partitions. In the training, each model was initialized with weights pre-trained on ImageNet \cite{deng2009imagenet}, and were fine-tuned on the internal hip OA dataset. The 4-fold cross-validations experiments were repeated 15 times to account for the random selections of the patients in the three partitions.  The models were further tested on the external dataset with the disease grade distribution shown in Table \ref{tb:datasets}. Specifically, a model trained on the entire internal dataset was used to predict the DRRs in the external testing dataset, and the predictions were evaluated independently.\\
\indent\textbf{Hyper-parameter settings.} The hyper-parameters of the grading models are shown in Table \ref{tb:subsec_hyper_parameter}. Classification and regression settings were used for each model, with 200 training epochs for classification and 300 epochs for regression.  In each fold,  the model with the highest accuracy on the validation partition was tested on the testing partition. The loss function was the focal loss \cite{Lin_2017_ICCV}  in the classification setting and the mean absolute error in the regression setting.  Both functions were minimized using Adam \cite{kingma2014adam} optimizer. The learning rate was adjusted dynamically using a Cosine Annealing scheduler.\\
\indent\textbf{Data augmentation.} Data augmentation was applied during training and inference using Albumentations (ver.1.1.0) \cite{info11020125}. The transformation parameters were set as follows: rotation (limit=15$^\circ$), blur (blur\_limit=(1,9)), contrast change (brightness\_limit=(-0.2,0.4), contrast\_limit=(-0.2,0.4)), masking (min\_holes=5, max\_holes=10) and intensity normalization (mean=[0.485,0.456,0.406], std=[0.229,0.224,0.225]) were used during training. 

\begin{table}[tb]
  \centering
  \caption{Hyper-parameter settings in training.}
  \scalebox{0.8}{
    \begin{tabular}{ccccc}
    \toprule
    Model       & Grading        & Epochs & Base LR          & Dropout rate \\
    \midrule
    ViT\_B16    & Classification & 200    & $5\times10^{-5}$ & 0.1\\
                & Regression     & 300    & $5\times10^{-5}$ & 0.1\\
    VGG16       & Classification & 200    & $5\times10^{-5}$ &  0.3 \\
                & Regression     & 300    & $8\times10^{-5}$ &  0.1 \\
    DenseNet161 & Classification & 200    & $5\times10^{-5}$ &  0.2 \\
                & Regression     & 300    & $8\times10^{-5}$ &  0.2 \\
    \bottomrule
    \end{tabular}
  }
  \label{tb:subsec_hyper_parameter}
\end{table}

\indent\textbf{Computation environment.} In this study, the experiments were implemented in Python using PyTorch framework (ver.0.12.0) \cite{paszke2017pytorch}, and model architectures were imported from Torchvision library (ver.1.11.0) \cite{torchvision2016}. The experiments were run on a linux-based GPU-cluster with nodes including the NVIDIA GPUs RTX2080ti (11GB),  RTX3090 (24GB) and RTX4090 (48GB).

\section{Results}
\label{sec:sec_results}

\begin{table}[tb]
  \caption{
    Summary of the exact class accuracy (ECA) and one-neighbor class accuracy (ONCA) obtained by the three models with separated and combined grading settings on the \textit{internal} dataset. The largest values between combined and separated settings are shown in \textbf{bold}, and the largest one in each row is additionally \textbf{\underline{underlined}}.
  }
  
  \centering
  \scalebox{0.64}{
    \begin{tabular}{@{}ccccc|ccc@{}}
    \toprule
    \multicolumn{8}{c}{\textbf{Exact class accuracy (Mean±SD)}}\\
    \midrule
    Number of samples & \multicolumn{4}{c|}{1 (w/o dropout)} & \multicolumn{3}{c}{50 (w/ dropout)} \\
    \midrule
    Model       & Grading        & Combined            & \multirow{2}{*}{\shortstack[l]{Separated\\Crowe, KL}}           & P-value & Combined            & \multirow{2}{*}{\shortstack[l]{Separated\\Crowe, KL}} & P-value \\
                &                &                     &                                                                 &         &                     &                     &         \\
    \midrule
    ViT\_B16    & Classification & \underline{\textbf{0.650}}±.029 & 0.638±.022          & n.s.    & \textbf{0.649}±.023 & 0.639±.023          & n.s. \\
                &                &                     & 0.926±.009, 0.713±.020          & \textbf{--}      &                     & 0.926±.009, 0.713±.024 & \textbf{--}   \\  
                & Regression     & 0.653±.016          & \textbf{0.658}±.014 & n.s.    & 0.656±.015          & \underline{\textbf{0.660}}±.010 & n.s. \\
                &                &                     & 0.919±.008, 0.739±.013          & \textbf{--}      &                     & 0.918±.007, 0.741±.011 & \textbf{--}   \\  
    VGG16       & Classification & 0.637±.020          & \underline{\textbf{0.643}}±.016 & n.s.    & 0.640±.017          & \textbf{0.642}±.021 & n.s. \\
                &                &                     & 0.924±.006, 0.719±.019          & \textbf{--}      &                     & 0.923±.007, 0.719±.018 & \textbf{--}   \\  
                & Regression     & 0.625±.029          & \underline{\textbf{0.657}}±.016          & n.s.    & 0.606±.028          & \underline{\textbf{0.656}}±.014 & *    \\
                &                &                     & 0.917±.005, 0.737±.014          & \textbf{--}      &                     & 0.917±.005, 0.738±.013 & \textbf{--}   \\  
    DenseNet161 & Classification & 0.634±.016          & \underline{\textbf{0.652}}±.015 & *       & 0.623±.022          & \textbf{0.632}±.020 & n.s. \\
                &                &                     & 0.922±.006, 0.730±.016          & \textbf{--}      &                     & 0.919±.008, 0.712±.019 & \textbf{--}   \\  
                & Regression     & 0.618±.015          & \underline{\textbf{0.663}}±.016 & *       & 0.587±.023          & \textbf{0.602}±.019 & n.s. \\
                &                &                     & 0.922±.007, 0.740±.014          & \textbf{--}      &                     & 0.896±.013, 0.700±.011 & \textbf{--}   \\  
    \bottomrule
    \\
    \toprule
    \multicolumn{8}{c}{\textbf{One-neighbor class accuracy (Mean±SD)}} \\
    \midrule
    Number of samples & \multicolumn{4}{c|}{1 (w/o dropout)} & \multicolumn{3}{c}{50 (w/ dropout)} \\
    \midrule
    Model       & Grading        & Combined            & Separated           & P-value & Combined            & Separated           & P-value \\
    \midrule
    ViT\_B16    & Classification & 0.958±.021          & \underline{\textbf{0.964}}±.028 & n.s.    & 0.955±.021          & \textbf{0.956}±.015 & n.s. \\
                & Regression     & 0.961±.010          & \textbf{0.964}±.012 & n.s.    & 0.962±.010          & \underline{\textbf{0.967}}±.012 & n.s. \\
    VGG16       & Classification & 0.948±.008          & \underline{\textbf{0.982}}±.005 & *       & 0.950±.007          & \underline{\textbf{0.982}}±.005 & *    \\
                & Regression     & \underline{\textbf{0.972}}±.004 & 0.969±.009          & n.s.    & \textbf{0.971}±.011 & 0.969±.008          & n.s. \\
    DenseNet161 & Classification & 0.953±.009          & \underline{\textbf{0.965}}±.009 & n.s.    & 0.947±.005          & \textbf{0.937}±.013 & *    \\
                & Regression     & 0.946±.009          & \underline{\textbf{0.961}}±.010 & *       & \textbf{0.932}±.007 & 0.927±.017          & *    \\
    \midrule
    \multicolumn{8}{l}{* Student's t-test between means of combined and separated settings (Bonferroni correction; P\textless2e-3).}\\
    \bottomrule
    \end{tabular}
  }
  
  \label{tb:subsec_1head2head_results}
\end{table}

\begin{table}[tb]
  \caption{
    Summary of accuracy and balanced accuracy for the exact class and one-neighbor class obtained by the three models with combined and separated grading settings on the \textit{external} dataset. The largest values between combined and separated settings are shown in \textbf{bold}, and the largest one in each row for each metric is additionally \textbf{\underline{underlined}}.
  }

  \begin{subtable}{\linewidth}
  \centering
  \scalebox{0.60}{
    \begin{tabular}{@{}cccccc|cccc@{}}
      \toprule
      \multicolumn{10}{c}{\textbf{Exact class}}\\
      \midrule
                        &     & \multicolumn{4}{c|}{Accuracy} & \multicolumn{4}{c}{Balanced accuracy}\\
      \midrule
      Number of samples &     & \multicolumn{2}{c}{1 (w/o dropout)} & \multicolumn{2}{c|}{50 (w/ dropout)} & \multicolumn{2}{c}{1 (w/o dropout)} & \multicolumn{2}{c}{50 (w/ dropout)} \\
      \midrule
      Model       & Grading & Combined & \multirow{2}{*}{\shortstack[l]{Separated\\Crowe, KL}} & Combined & \multirow{2}{*}{\shortstack[l]{Separated\\Crowe, KL}} & Combined & \multirow{2}{*}{\shortstack[l]{Separated\\Crowe, KL}} & Combined & \multirow{2}{*}{\shortstack[l]{Separated\\Crowe, KL}} \\
                  &                 &          &                    &          &                                &                               &           &                   &      \\
      \midrule
      ViT\_B16    & Classification  & 0.519    & \textbf{0.529}     & 0.481    & \underline{\textbf{0.558}}     & \underline{\textbf{0.498}}    & 0.437     & \textbf{0.479}    & 0.467 \\
                  &                 &          &  0.817, 0.712      &          & 0.827, 0.731                   &                   & 0.503, 0.570          &        &  0.528, 0.592       \\
                  & Regression      & \underline{\textbf{0.567}}    & \underline{\textbf{0.567}}     & \textbf{0.538}    & \textbf{0.538}     & 0.474    & \underline{\textbf{0.475}}     & 0.437    & \textbf{0.454} \\
                  &                 &          & 0.817, 0.731       &          & 0.827, 0.692                   &                               & 0.505, 0.599       &                   & 0.533, 0.542    \\
      VGG16       & Classification  & 0.529    & \underline{\textbf{0.588}}     & 0.538    & \textbf{0.587}     & 0.428    & \textbf{0.460}     & 0.435    & \underline{\textbf{0.487}} \\
                  &                 &          & 0.788, 0.760             &          &  0.788, 0.788            &                               & 0.435, 0.669           &      & 0.435, 0.710       \\
                  & Regression      & 0.500    & \textbf{0.519}     & \underline{\textbf{0.538}}    & 0.529     & 0.395    & \textbf{0.419}     & \underline{\textbf{0.436}}    & 0.429 \\
                  &                 &          & 0.798, 0.721             &          & 0.798, 0.731             &                          & 0.445, 0.571          &                   & 0.445, 0.589      \\
      DenseNet161 & Classification  & 0.481    & \underline{\textbf{0.558}}     & \textbf{0.481 }   & 0.471     & 0.373    & \underline{\textbf{0.500}}     & 0.353    & \textbf{0.355} \\
                  &                 &          &  0.837, 0.721      &          &  0.750,702                     &                               & 0.545, 0.613      &                   & 0.275, 0.618     \\
                  & Regression      & 0.558    & \underline{\textbf{0.606}}     & \textbf{0.548}    & \textbf{0.548}     & 0.476    & \underline{\textbf{0.522}}     & 0.442    & \textbf{0.453} \\
                  &                 &          & 0.808, 0.798                  &          & 0.788, 0.740                      &                               & 0.516, 0.684         &                   & 0.441, 0.613   \\
      \bottomrule
    \end{tabular}
  }
  \end{subtable}
  \\
  \\
  \begin{subtable}{\linewidth}
  \centering
    \scalebox{0.63}{
      \begin{tabular}{@{}cccccc|cccc@{}}
      \toprule
      \multicolumn{10}{c}{\textbf{One-neighbor class}}\\
      \midrule
                        &     & \multicolumn{4}{c|}{Accuracy} & \multicolumn{4}{c}{Balanced accuracy}\\
      \midrule
      Number of samples &     & \multicolumn{2}{c}{1 (w/o dropout)} & \multicolumn{2}{c|}{50 (w/ dropout)} & \multicolumn{2}{c}{1 (w/o dropout)} & \multicolumn{2}{c}{50 (w/ dropout)} \\
      \midrule
      Model       & Grading         & Combined & Separated & Combined & Separated & Combined & Separated & Combined & Separated \\
      \midrule
      ViT\_B16    & Classification  & 0.904    & \textbf{0.923}     & 0.913    & \underline{\textbf{0.942}}     & 0.878    & \textbf{0.898}     & 0.892    & \underline{\textbf{0.923}} \\
                  & Regression      & \underline{\textbf{0.913}}    & 0.894     & \textbf{0.904}    & 0.894     & \underline{\textbf{0.888}}    & 0.860     & \textbf{0.866}    & \textbf{0.866} \\
      VGG16       & Classification  & 0.885    & \underline{\textbf{0.942}}     & 0.894    & \underline{\textbf{0.942}}     & 0.838    & \underline{\textbf{0.939}}     & 0.853    & \underline{\textbf{0.939}} \\
                  & Regression      & \underline{\textbf{0.965}}    & 0.923     & \textbf{0.923}    & \textbf{0.923}     & 0.795    & \textbf{0.890}     & \underline{\textbf{0.891}}    & 0.890 \\
      DenseNet161 & Classification  & 0.885    & \underline{\textbf{0.942}}     & 0.798    & \textbf{0.808}     & 0.847    & \underline{\textbf{0.931}}     & 0.694    & \textbf{0.713} \\
                  & Regression      & \underline{\textbf{0.952}}    & 0.942     & 0.798    & \textbf{0.808}     & \underline{\textbf{0.941}}    & 0.931     & \textbf{0.758}    & 0.745 \\
      \bottomrule
    \end{tabular}
    }
  \end{subtable}
  
  \label{tb:subsec_external_results}
\end{table}

\subsection{Grading accuracy }
\label{sec:subsec_grading_accuracy}
\begin{figure}[H]
  \centering
  \subcaptionbox{Combined, 50 samples}{%
    \includegraphics[width=0.5\textwidth]{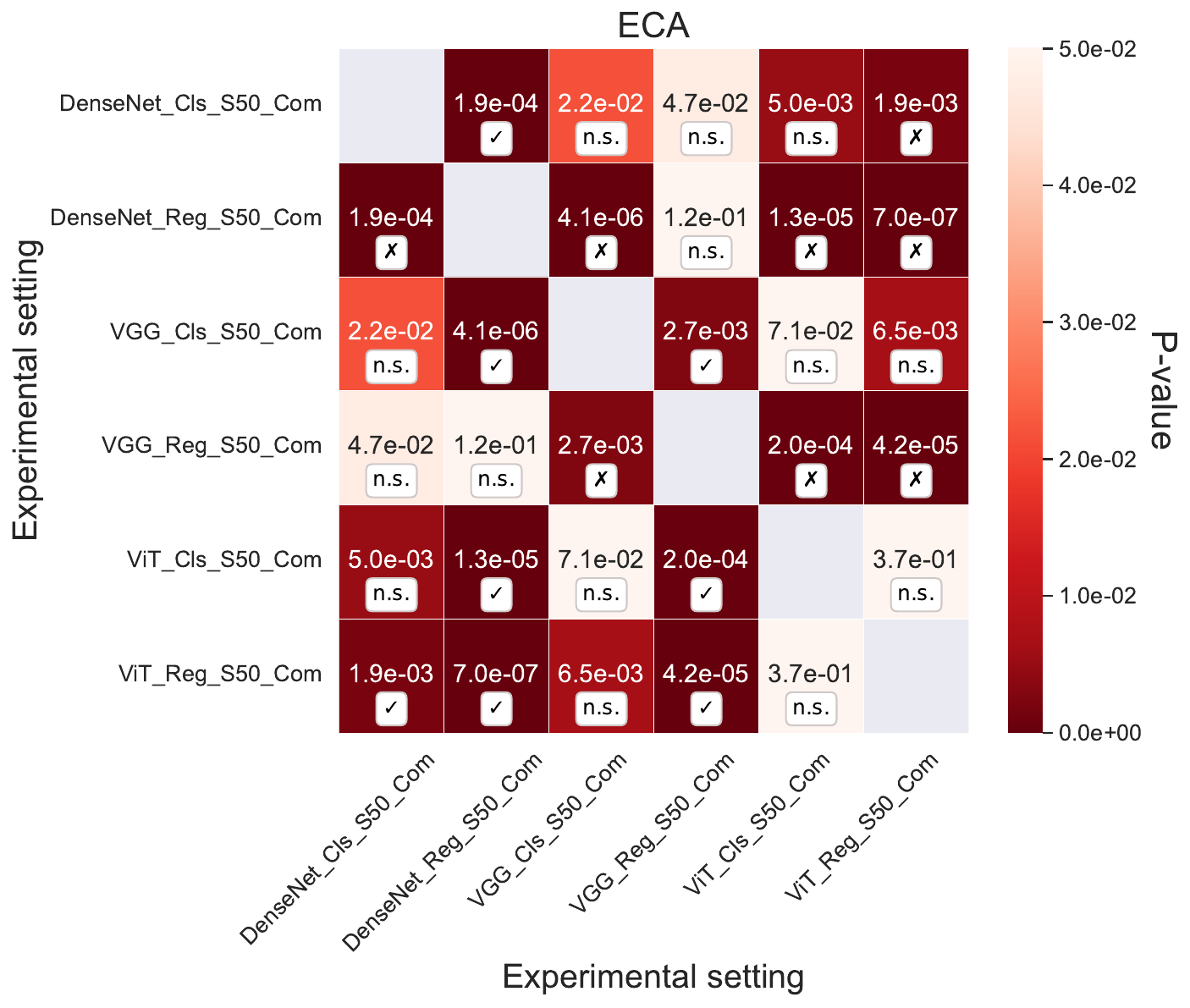}%
    \label{subfig_a}%
  }%
  \subcaptionbox{Separated, 50 samples}{%
    \includegraphics[width=0.5\textwidth]{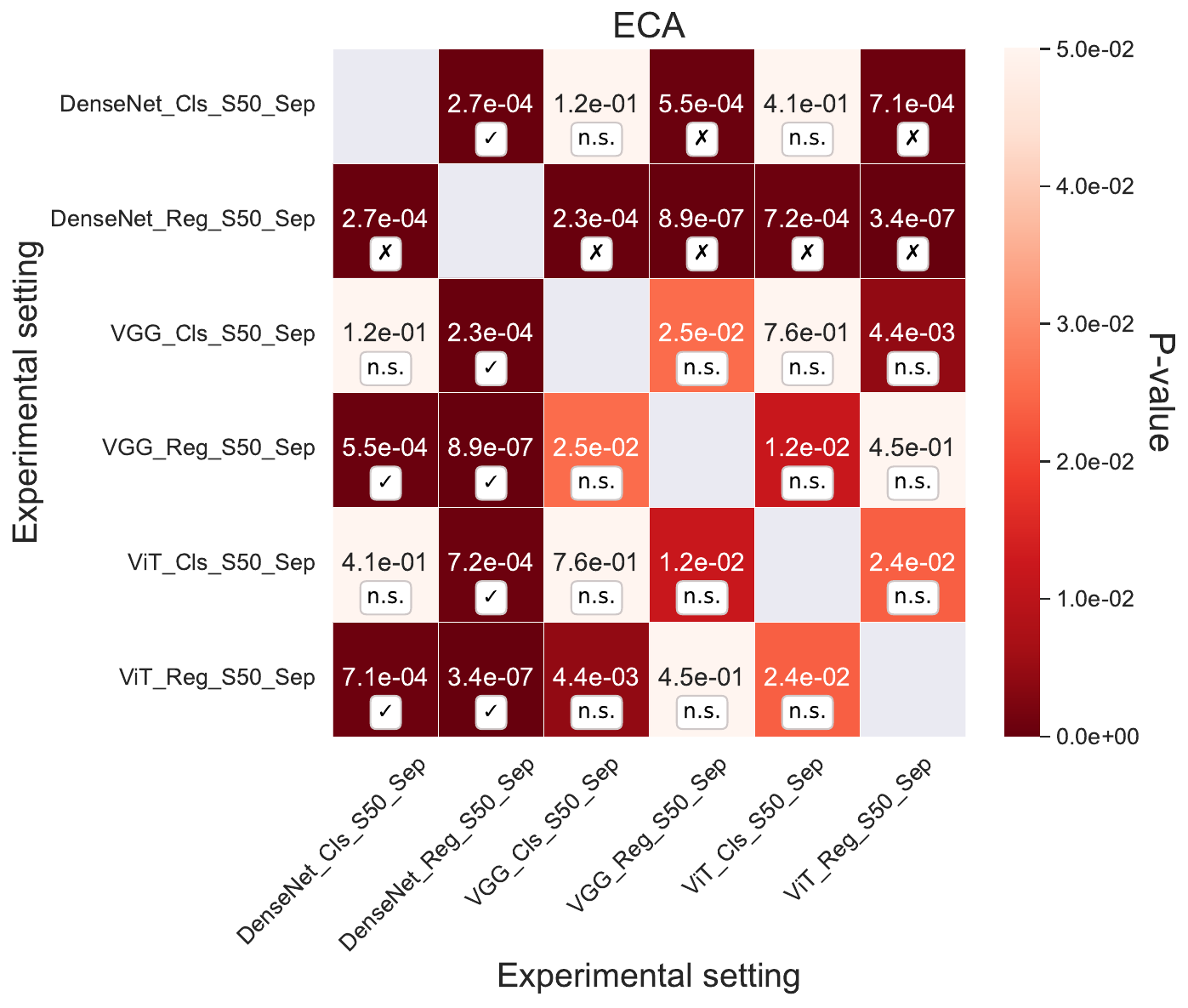}%
    \label{subfig_b}%
  }%
  \\
    \subcaptionbox{\hspace{0.6cm}Combined \newline \parbox{\linewidth}{\centering 1 vs 50 samples}}{%
    \includegraphics[width=0.30\textwidth]{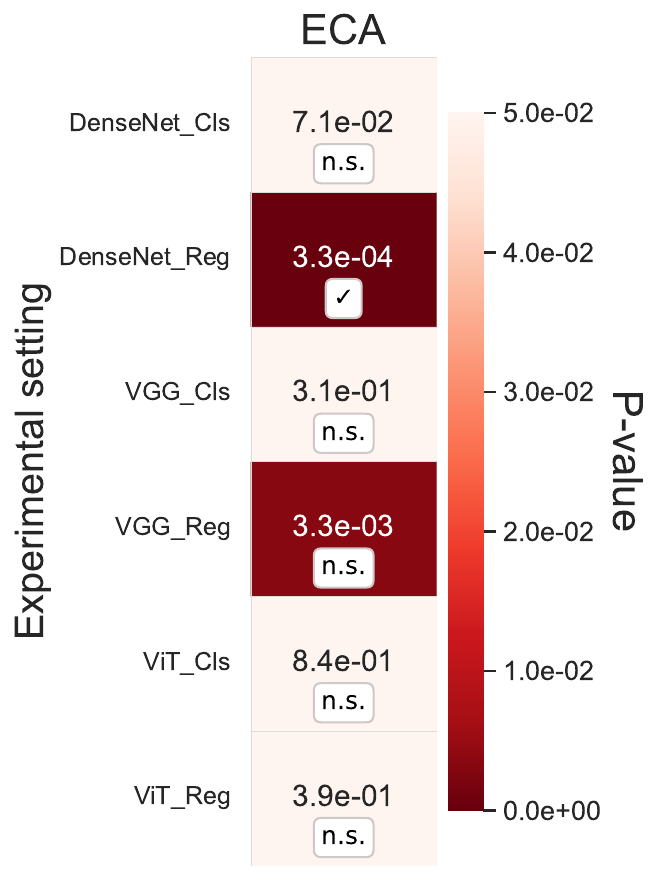}%
    \label{subfig_c}%
  }%
  \hspace{1.7cm}
  \subcaptionbox{\hspace{0.6cm}Separated \newline \parbox{\linewidth}{\centering 1 vs 50 samples}}{%
    \includegraphics[width=0.30\textwidth]{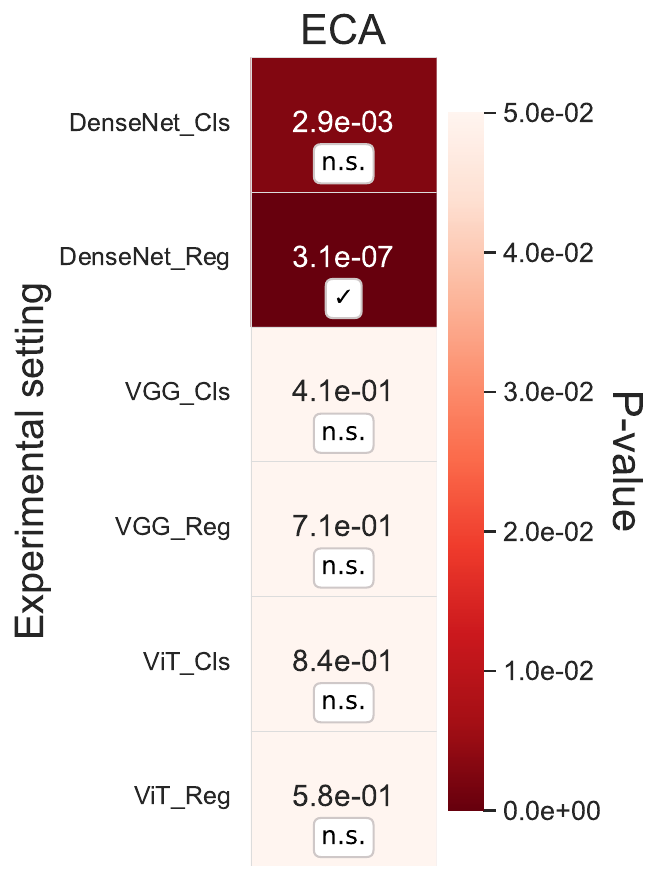}%
    \label{subfig_d}%
  }%

  \caption{P-values of the differences between the ECA of the three models and the prediction methods under the combined and separated label as well as 1 and 50 samples settings. \textbf{\checkmark} in (a, b) indicates that the vertical experimental settings had higher accuracy than the horizontal setting; for (c, d), sample 1 setting had higher accuracy than the samples 50 setting with a statistically significant difference (Student's t-test with Bonferroni correction, P\textless3e-3 for (a, b), P\textless1e-3 for (c, d)). \textbf{\ding{55}} in (a, b) indicates that the vertical settings yielded lower accuracy;  for (c, d), sample 1 yielded lower accuracy with a statistically significant difference, while \textbf{n.s.} indicates no significant difference was observed. Reg: regression, Cls: classification, S50: 50 samples (w/ dropout), Com: combined, Sep: separated.}
  \label{fig:figure03}
\end{figure}

\begin{figure}[tb]
  \centering
  \includegraphics[width=12cm]{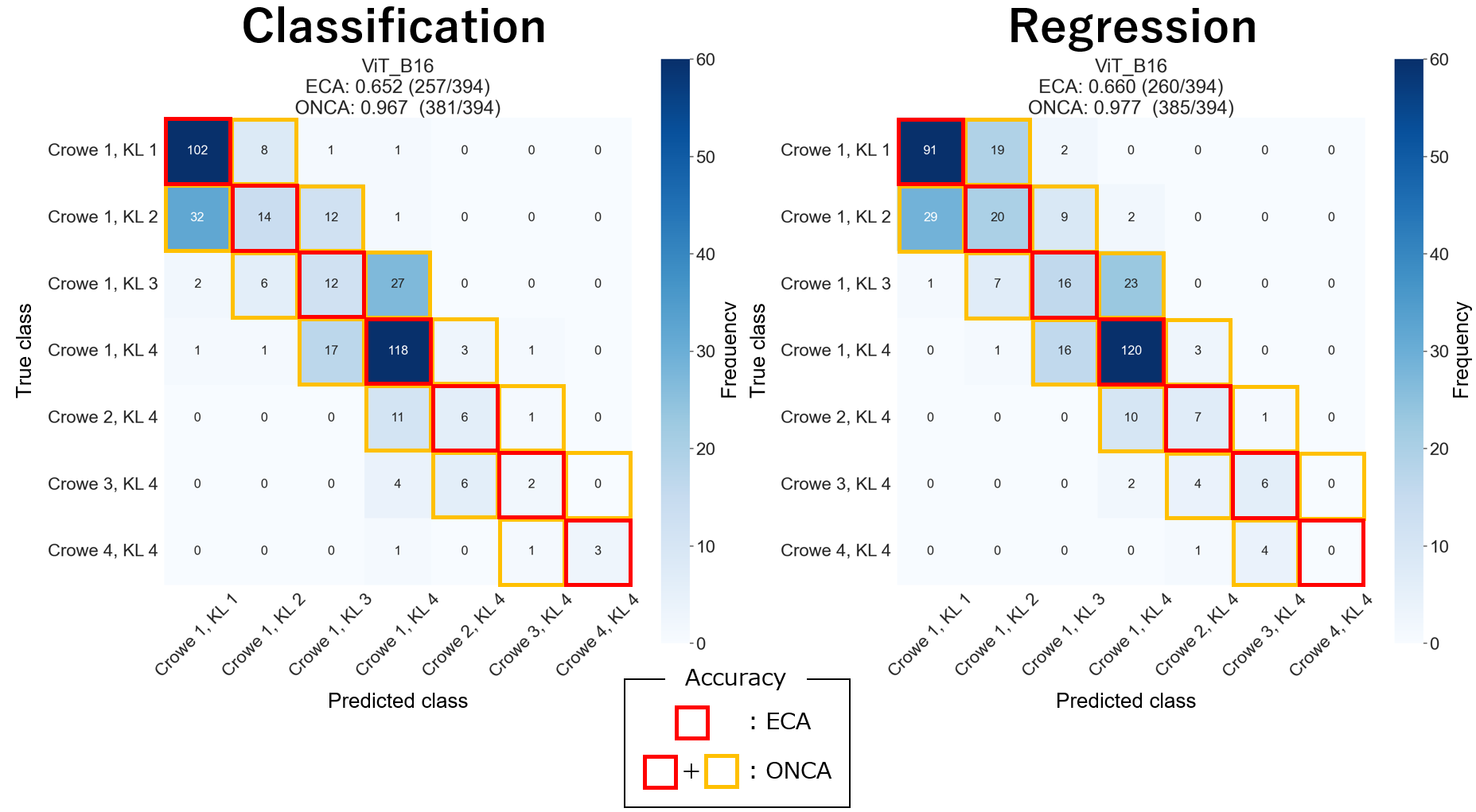}
  \caption{
    Confusion matrices of the ViT grading model in classification (left) and regression (right) settings. The confusion matrices correspond to the models yielding median ECA in both settings.
  }
  \label{fig:figure04}
\end{figure}

\indent\indent\textbf{Internal dataset.}  Table \ref{tb:subsec_1head2head_results} shows the overall accuracy of the three models for 1 sample (w/o dropout) and 50 samples (w/ dropout) experiments in the combined and separated label predictions. Figure \ref{fig:figure03} shows the p-values of the differences between the models at the different configurations. The highest ECA was obtained under the separated and regression settings using ViT (0.660±0.010) and DenseNet (0.663±0.016) models, while the ONCAs were $>$0.90 in all models. In the combined setting with 50 samples, ViT's regression significantly outperformed the other methods in ECA (0.656±0.015; See Figure \ref{fig:figure03} (a)). In the separated setting, while ViT's regression showed the highest ECA, there was no significant difference from that of VGG's regression (0.656±0.014), suggesting that both are similarly superior to other methods (See Figure \ref{fig:figure03} (b)). In comparing combined and separated settings, VGG's regression showed a statistically significant improvement in the separated setting, while ViT did not. Crowe grading has shown larger ECA than KL in all settings, where the largest Crowe and KL accuracy was shown using ViT model in the classification and regression settings, respectively. The results of t-tests for conditions other than those mentioned in Fig. \ref{fig:figure03} are shown in Appendix B Fig. \ref{fig:appendix_pvalue}. The confusion matrices in Fig. \ref{fig:figure04} correspond to the repetitions that yielded the median accuracy under the classification and combined settings with 50 samples. The ECA of low-severity cases was high, whereas it was lower for high-severity grades (Crowe$\ge$2, KL 4). ONCA was higher in regression than in classification settings in high-severity cases.

\begin{figure}[tb]
  \centering
  \includegraphics[width=7cm]{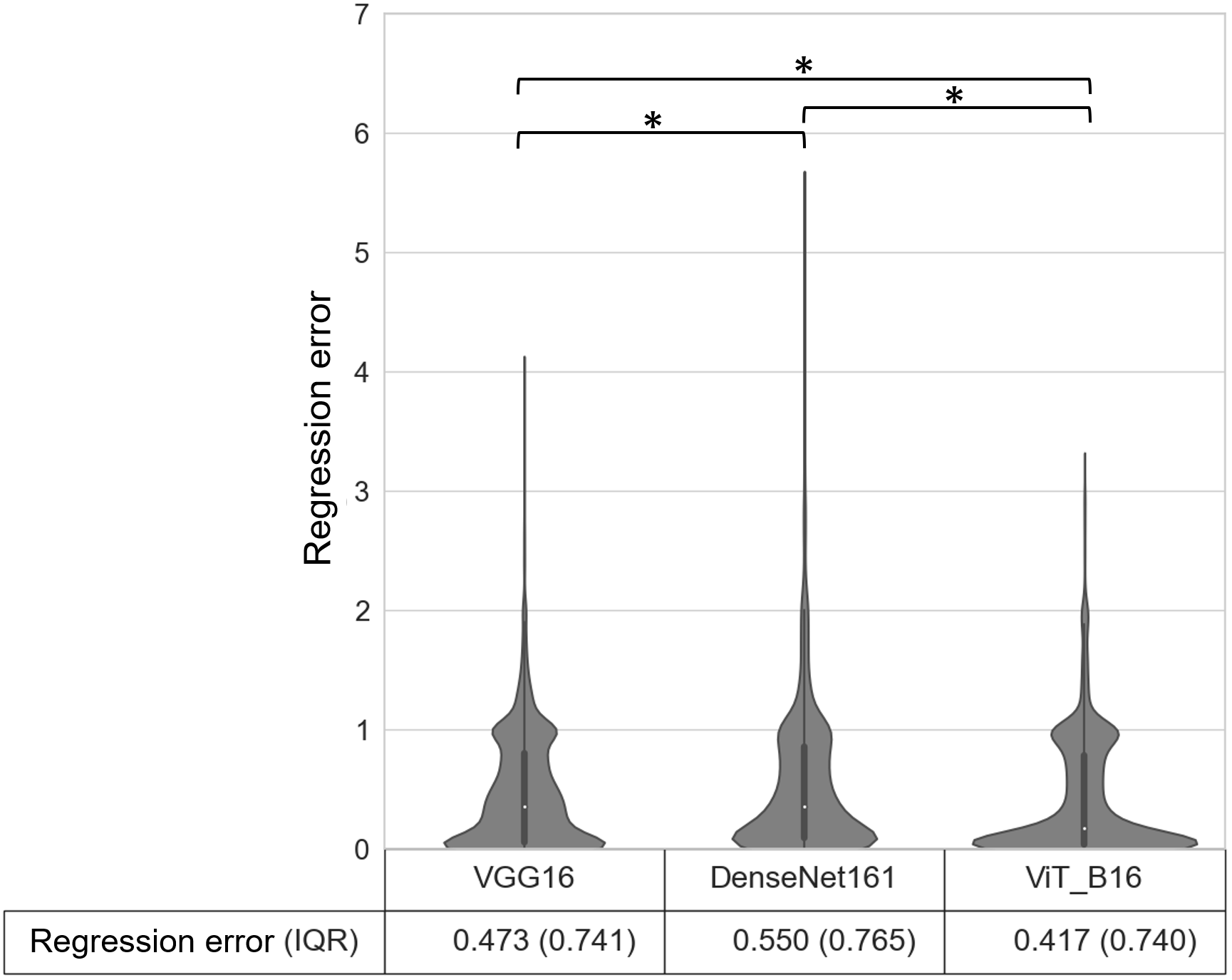}
  \caption{
    Distributions of the regression errors in each model under the combined setting. The table shows mean regression errors with inter-quartile ranges (IQR) (Mann-Whitney's U-test; Bonferroni correction P\textless0.02). 
  }
  \label{fig:figure05}
\end{figure}

Figure \ref{fig:figure05} shows the regression errors (difference between true and predicted classes in regression and combined settings) and their distributions by the grading models. ViT produced the smallest error, which was 0.383 (0.670 IQR: inter-quartile range). The three models had comparable IQRs. Statistically significant differences were obtained between ViT and the other models (Mann-Whitney's U-test, Bonferroni correction, P\textless0.02).

\begin{figure}[tb]
  \centering
  \includegraphics[width=8cm]{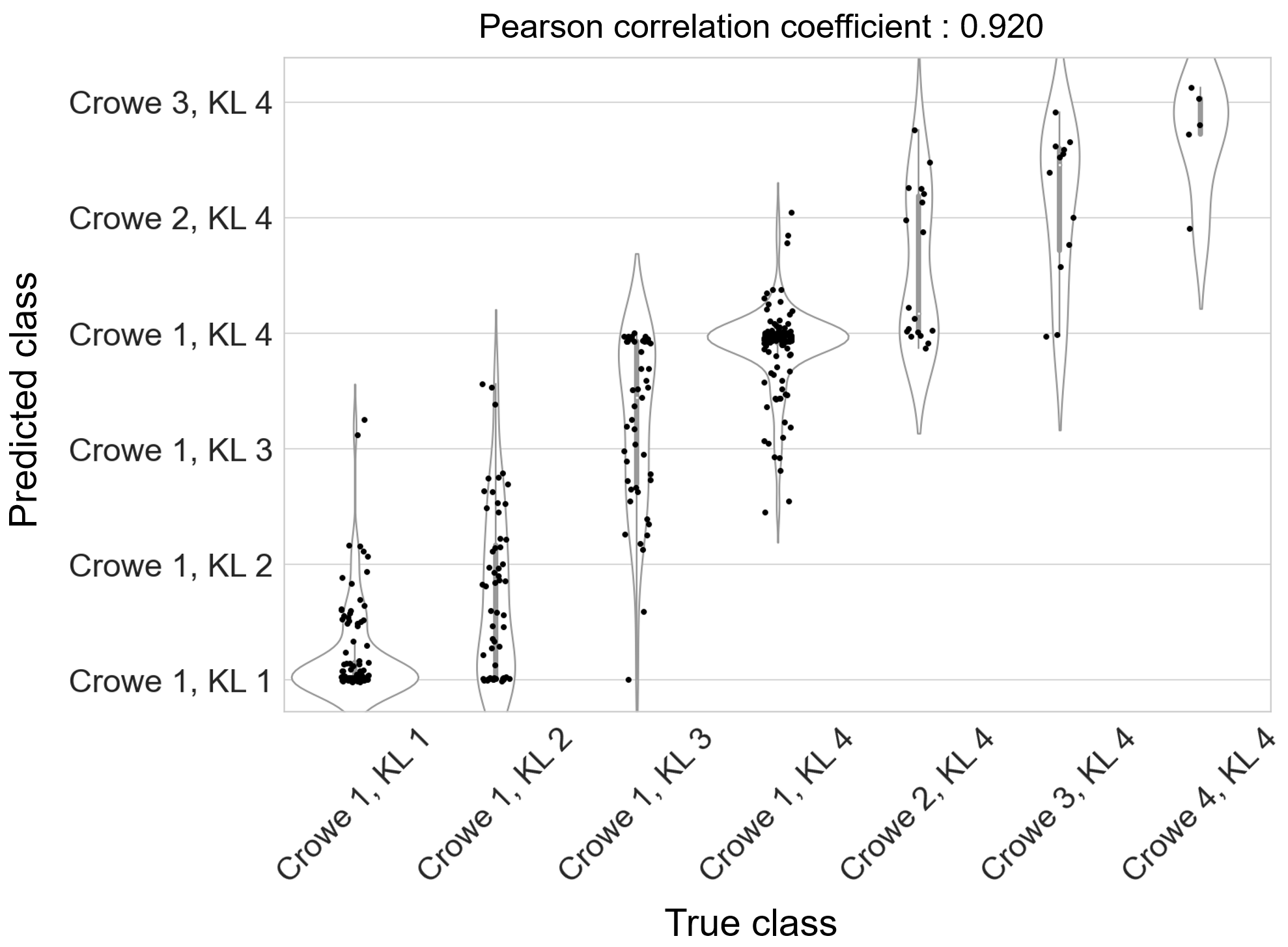}
  \caption{
    Relationship between the true and predicted classes by the ViT model in the regression and combined settings in the experiment corresponding with the median ECA.
  }
  \label{fig:figure06}
\end{figure}

\indent Figure \ref{fig:figure06} shows the relationship between the true and predicted classes by the ViT model in the regression and combined settings at the 15 cross-validation experiments. A positive strong correlation (Pearson correlation coefficient=0.920) was observed, indicating that the model could, to a large extent, adequately learn the continuous progression of the disease.

\begin{figure}[tb]
  \centering
  \includegraphics[width=7cm]{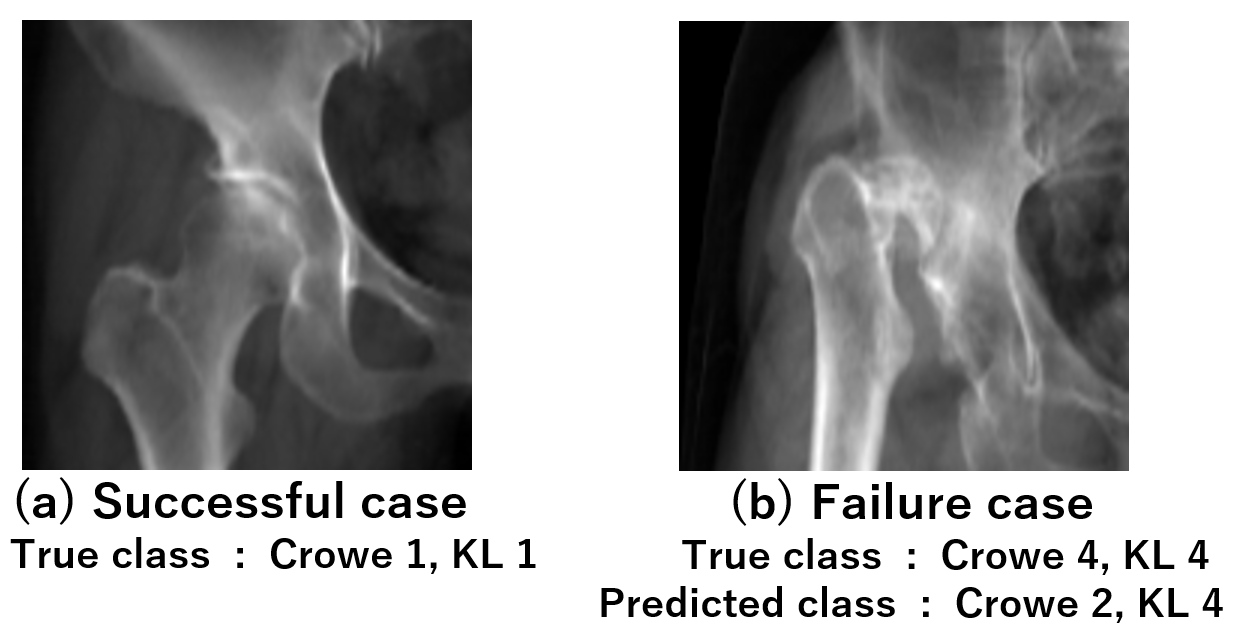}
  \caption{
    Representative cases about the performance of ViT. (a) Successful case,  (b) Failure case.
  }
  \label{fig:figure07}
\end{figure}

Figure \ref{fig:figure07} shows representative cases for successful and failure cases in the regression and combined settings of the ViT model. Figure \ref{fig:figure07}(a) shows a normal hip that was correctly classified as (Crowe 1, KL 1). In contrast, Figure \ref{fig:figure07}(b) shows a high severity hip (Crowe 4, KL 4) that was classified as (Crowe 2, KL 4).

\indent\textbf{External dataset.}
Table \ref{tb:subsec_external_results} shows the results of the external testing dataset (See Table \ref{tb:datasets}). The accuracy was overall lower than that of the internal dataset. This was caused by the difference in the distribution of grades in the two datasets. Specifically, the external dataset's proportion of severe cases falling into classes 5, 6, and 7 is, on average, about 6\% larger than the internal one. That made it more apparent that the model could not capture the characteristics of severe cases well due to a lack of training data. 
Therefore, for the external dataset, balanced accuracy, a metric more suitable for handling class imbalances, was employed.
The highest exact class balanced accuracy was obtained from DenseNet with 0.522 under 1 sample and separated setting. However, when the dropout sample was set to 50, the accuracy dropped remarkably. This shows possible dependency in DenseNet's performance on the implemented dropout layer configuration \cite{wan2019reconciling}. In all experiments, the balanced accuracy of KL was higher than that of Crowe, which emphasized the dependency in the overall performance on the small number of severe Crowe classes.

\subsection{Uncertainty analysis}
\label{sec:subsec_uncertainty_analysis}
\begin{figure}[tb]
  \centering
  \includegraphics[width=8cm]{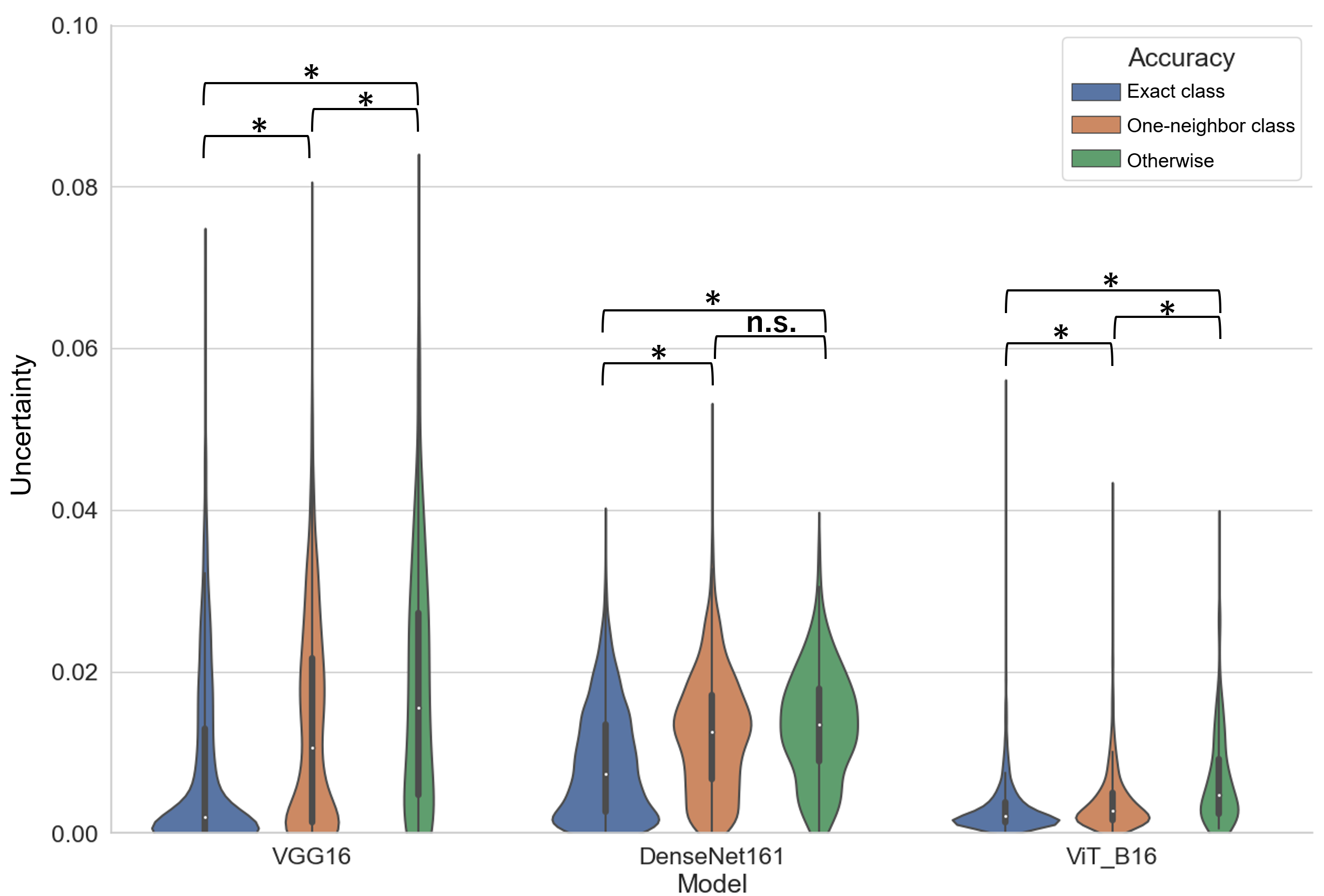}
  \caption{
    Analysis of the estimated uncertainty for the predictions of the three models in terms of classification accuracy (Mann-Whitney U-Test;  Bonferroni correction P\textless6e-3).
  }
  \label{fig:figure08}
\end{figure}

Figure \ref{fig:figure08} shows the uncertainty (variance of softmax probabilities shown in Eq. \ref{eq:uncertainty_equation}) of the three models in the classification and combined settings. Notably, cases corresponding to the exact class accuracy (blue) had relatively lower uncertainty, thus showing high model confidence. On the other hand, misclassified cases with large errors (green) had higher uncertainty. Statistically significant differences between the groups of large-error cases and correctly classified (exact class accuracy) ones were obtained (Mann-Whitney's U-test; Bonferroni correction, P$<$6e-3). ViT produced the lowest uncertainty levels among the three models.

\subsection{Learned representations}
\label{sec:subsec_learned_representations}
\begin{figure*}[tb]
  \centering
  \includegraphics[width=12cm]{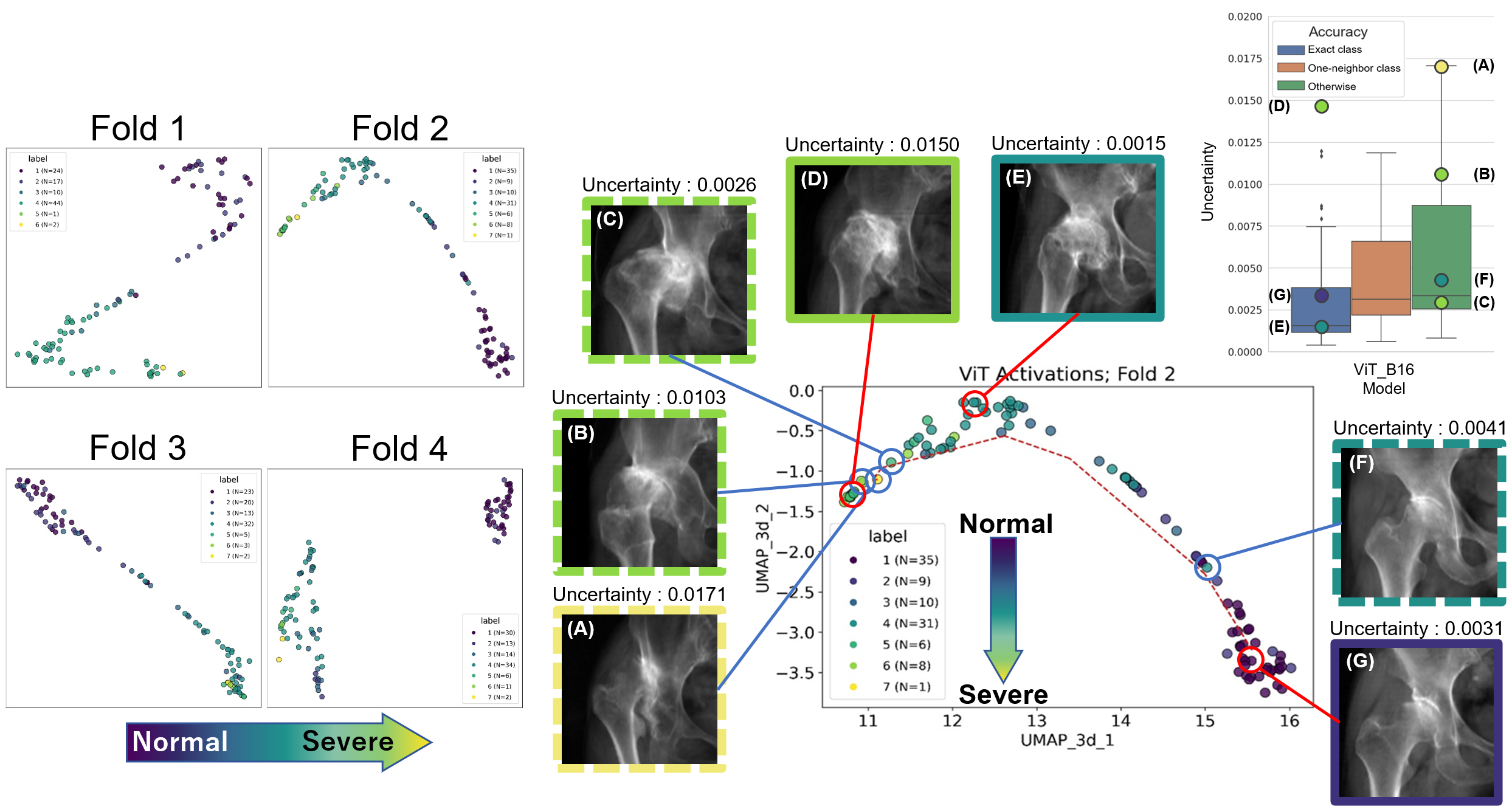}
  \caption{
    Analysis of the relationship between learned representations by the ViT model and the disease progression. \textbf{Left}: Feature map visualization of the four folds. \textbf{Right}: Enlarged feature map with representative cases from Fold 2 with their uncertainty in the upper right boxplot. Solid lines indicate successful cases, and dashed lines indicate failure ones.
  }
  \label{fig:figure09}
\end{figure*}

To analyze the relationship between the learned representations by the ViT model and disease progression, a Uniform Manifold Approximation and Projection (UMAP) \cite{mcinnes2018uniform} analysis was applied to the ViT feature vectors obtained in the classification and combined settings from each DRR. The feature vectors were obtained from the fully-connected layer before the output layer. 
Figure \ref{fig:figure09} shows 2D projections of the feature vectors in the UMAP space obtained from the model that produced the median ECA. For instance, the scatter plot of Fold 2 shows sequentially distributed dots (each of which represents a DRR image) w.r.t disease severity, from purple (normal) to yellow (severe). A similar pattern can be observed in Folds 1 and 3 plots, and the separation between low and high classes is apparent in Fold 4. This indicates the model's capability to capture the representative variations accompanying the disease progression.

In Fig. \ref{fig:figure09}, DRR images of representative cases were shown in the colored frames (solid lines for successful cases and dashed lines for failed ones under the combined setting). The successful examples clearly showed increased stages of the disease progression, where in case G (low severity) the femoral head was covered by the acetabulum, while in case D, the femoral head was clearly dislocated. However, in the failure examples, the model mistakenly classified case A as having a lower severity (Crowe 1, KL 4), despite its high severity (Crowe 4, KL 4). Case F (Crowe 1, KL 4) was predicted as a less severe class (Crowe 1, KL 2). When consulting with the orthopedic surgeon who annotated the dataset, he confirmed that the original GT annotation was incorrect in this case, and outweighed the model prediction of lower severity (Crowe 1, KL 2). The upper right plot in Fig. \ref{fig:figure09} shows the uncertainty distribution corresponding to the scatter plots. Misclassified cases had a higher uncertainty than the correctly classified ones. 

\section{Discussion}
\label{sec:sec_discussion}
In this study, an automated grading approach representing the disease progression of hip OA was proposed. The study is the first to validate multiple deep learning models under different settings, including combined and separated labeling based on Crowe and KL grades. The study has shown high ONCA ($>$0.90) in all models, which facilitates automated grading in large-scale CT databases and indicates the potential for further disease progression analysis. Given that conventional X-ray imaging is the gold-standard for diagnosis of hip OA \cite{ureten2020detection, von2020development}, we will consider the validation of our method to conventional X-ray images in our future work.
Subtle differences were observed between the 1 and 50 samples in the external dataset, with a common trend of degraded performance in the severe Crowe classes. For example, the model trained in regression and combined settings showed high accuracy in classifying normal to mild stages, as shown in Fig. \ref{fig:figure07}(a). However, the model showed lower accuracy in classifying severe cases, as shown in Fig. \ref{fig:figure07}(b). A similar trend was observed from Fig. \ref{fig:figure06}. Additionally, the study revealed that cases with classification errors had a higher uncertainty than the correctly classified ones. This indicates the possibility of using model uncertainty as a surrogate for hip OA classification accuracy and error detection.

Separated setting is theoretically capable of handling cases such as Crowe 2, KL 1. However, in the combined setting, even if the model could learn the features of Crowe 4 and KL 1 grades, it would predict it as one from within the trained labels. This shows the benefit of the separated prediction scheme in learning all possible combinations.

In the regression models using the combined and separated settings, DenseNet showed significantly lower accuracy when the number of dropout samples was set to 50, possibly caused by insufficient configuration, i.e., impeded feature-reuse in the layers after the dropout layer \cite{wan2019reconciling}. From Fig. \ref{fig:figure03} (c, d), it can be confirmed that ViT and VGG showed no significant difference between 1 sample and 50 samples.

As shown in Table 5, the balanced accuracy used in the external dataset is approximately 10\% lower than the unbalanced accuracy for both the exact and one-neighbor classes. The balanced accuracy was less affected by the accuracy of the majority classes 1 and 4.
In other words, this result shows limited accuracy in the severe classes, which have a smaller number of cases in the internal dataset.

In previous studies, hip OA was classified with an accuracy of 80--90\% \cite{gebre2022detecting,von2020development}; however, hip OA was treated as a binary classification problem, which may not represent the disease progression captured by our study. Indeed, the UMAP analysis in Fig. \ref{fig:figure09} showed that ViT model can capture the variability associated with the disease progression. We consider that the combined class of Crowe and KL grade successfully represented the disease progression.

As a limitation, our grading models showed lower accuracy in classifying high-severity cases. It is noteworthy that those cases are usually easy to grade by human experts due to the clear signs of deformed joints. One reason for the lower accuracy in those cases could be a small number and large variations of severe cases in this study (N(Class 5,6,7) = 35(9\%)).  This tendency was confirmed in the external testing dataset. The internal dataset used for training the final model was small in scale, limiting the performance on the external dataset.  To solve this problem, we plan to largely increase the cases from those classes. We have more than 2000 CT scans of hip OA that have not been graded, and we are considering applying the automatic grading developed in this study to that data. The cases with high severity and uncertainty will be detected by our method and will be assigned ground-truth labels by the medical experts. This will help to efficiently expand the training data and improve the grading accuracy of severe cases. Furthermore, the study was validated on CT images collected from a single institution. Future work will include experiments using datasets obtained by the CT scanners of different manufacturers and models. Another limitation is that we have graded the data into seven classes according to the Crowe and KL distributions in our database. However, there is a possibility of cases diagnosed with large Crowe and low KL grades, which did not exist in the current database. This could be addressed by extending the assigned classes to cover more possible grade combinations.  In addition, this study combined KL grades 0 and 1 into a single class due to the difficulty in distinguishing between healthy hip joints and early-stage OA \cite{gebre2022detecting}. The automated classification of the two classes would further help in the early detection of hip OA, thus potentially allowing for the proposal of appropriate treatment strategies to prevent progression. 
\\


\section{Conclusion}
\label{sec:sec_conclusion}
In this study, we proposed an automated method grading hip OA in DRRs derived from CT images. The study investigated the usability of three deep learning models for predicting Crowe and KL grades under several labeling and inference settings. The models showed particularly high ONCA, which facilitates automated grading in large-scale CT databases and indicates the potential for further disease progression analysis.  Furthermore, the study has shown the potential of model uncertainty as a surrogate of hip OA classification accuracy. 

\backmatter

\bmhead{Acknowledgments}

This work was funded by MEXT/JSPS KAKENHI (19H01176, 20H04550, 21K16655, 21K18080).

\section*{Declarations}
\textbf{Conflict of interest} Nothing to declare.

\noindent\textbf{Ethics approval} Ethical approval was obtained from the Institutional Review Boards (IRBs) of the institutions participating in this study (IRB approval numbers: 21115 for Osaka University Hospital and 2020-M-7 for Nara Institute of Science and Technology.)

\bibliographystyle{unsrtnat}

\clearpage
\appendix
\begin{appendices}



\section{Dropout rate}
\begin{figure}[h]
  \centering
  \begin{subfigure}{.45\linewidth}
    \centering
    \includegraphics[width=\linewidth]{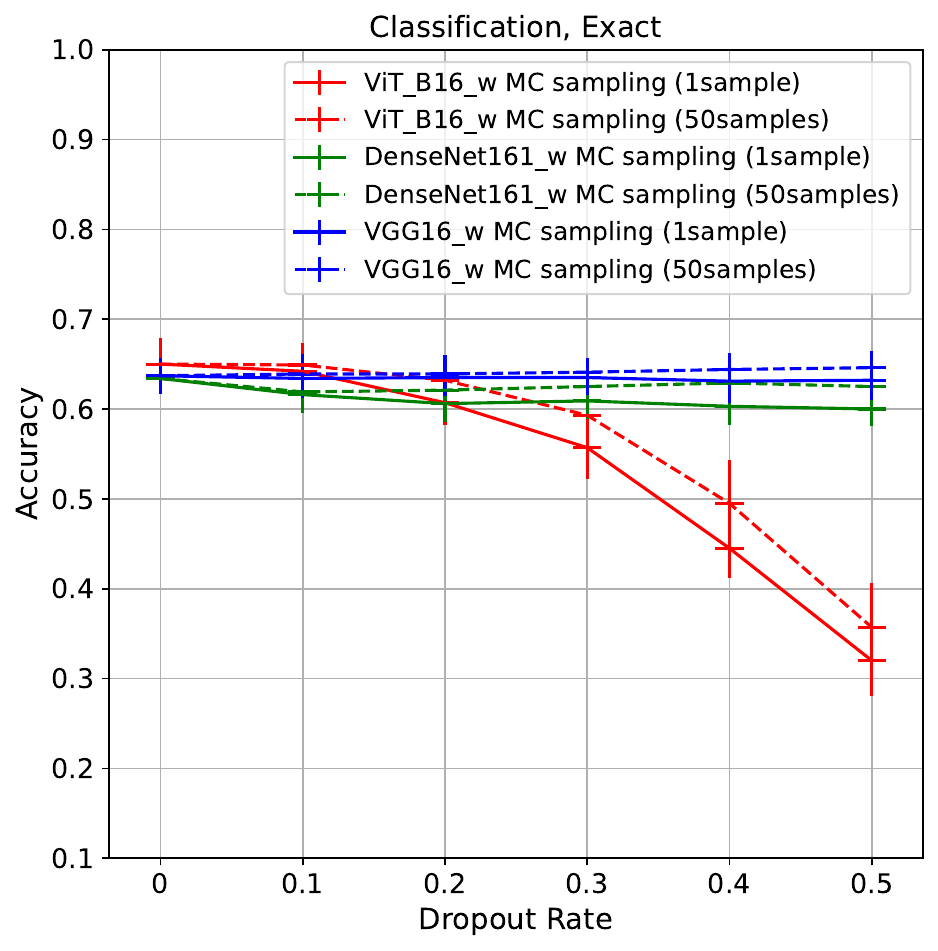}
    \label{fig:subfig_dropout_1}
  \end{subfigure}%
  \begin{subfigure}{.45\linewidth}
    \centering
    \includegraphics[width=\linewidth]{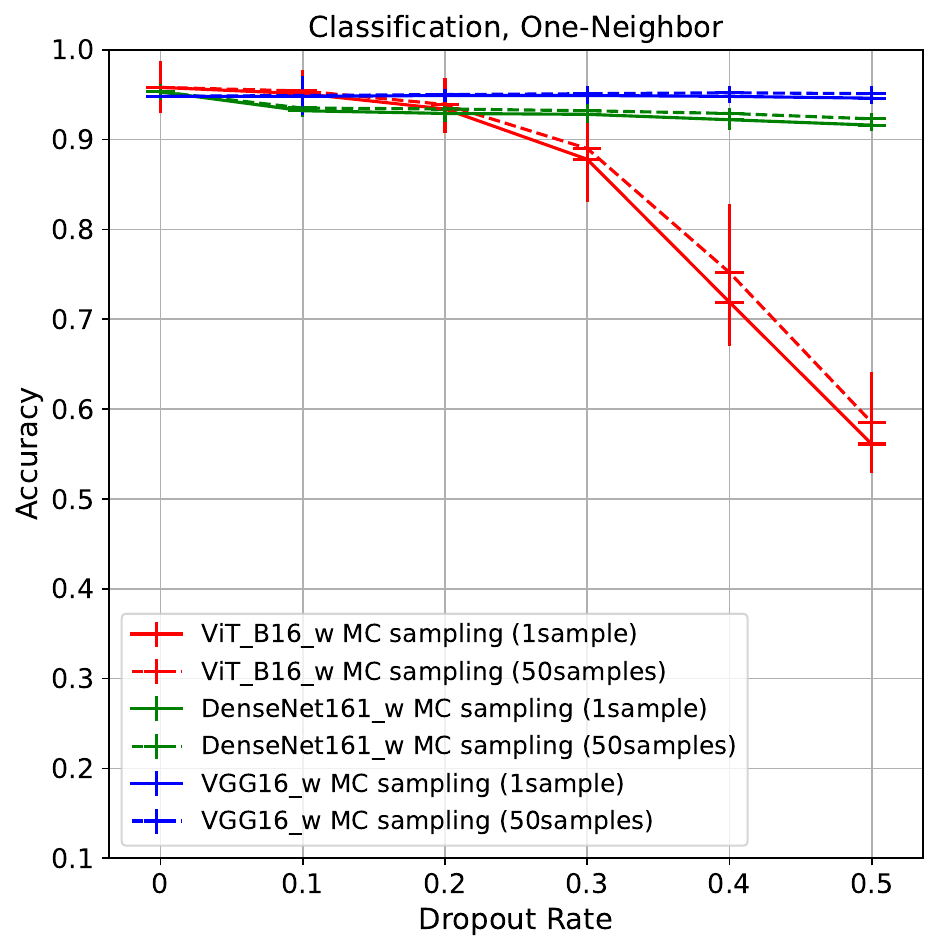}
    \label{fig:subfig_dropout_2}
  \end{subfigure}
  \begin{subfigure}{.45\linewidth}
    \centering
    \includegraphics[width=\linewidth]{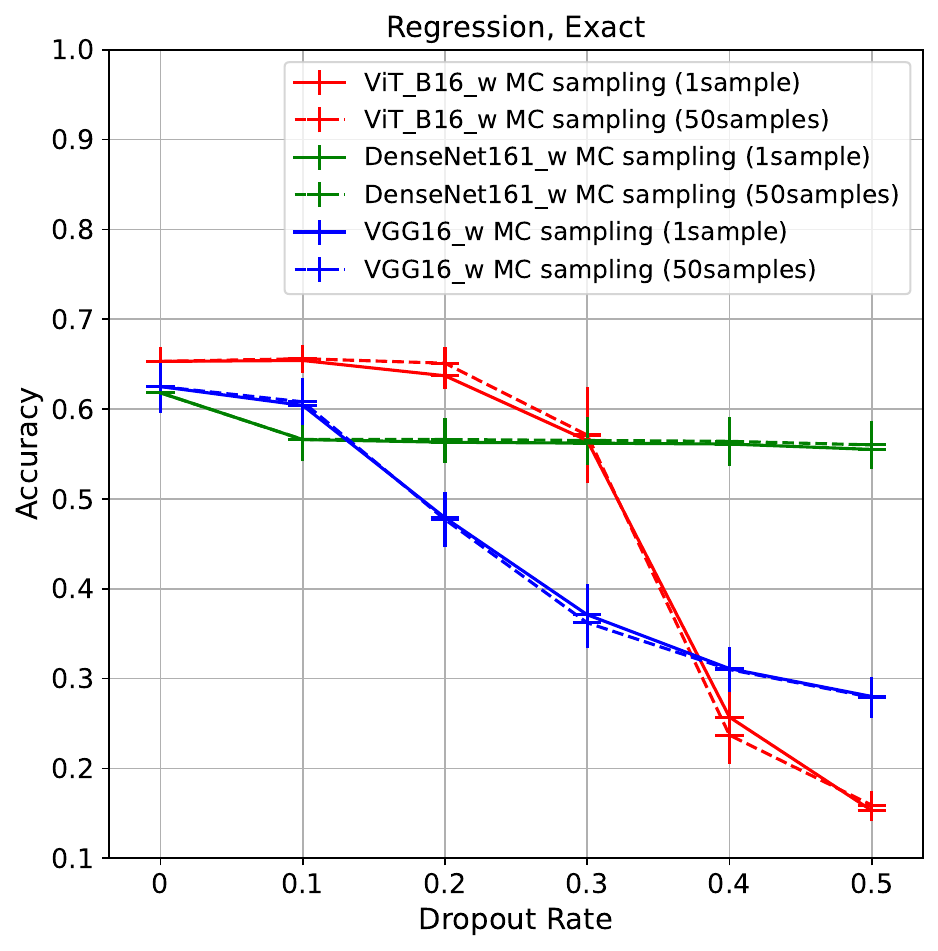}
    \label{fig:subfig_dropout_3}
  \end{subfigure}%
  \begin{subfigure}{.45\linewidth}
    \centering
    \includegraphics[width=\linewidth]{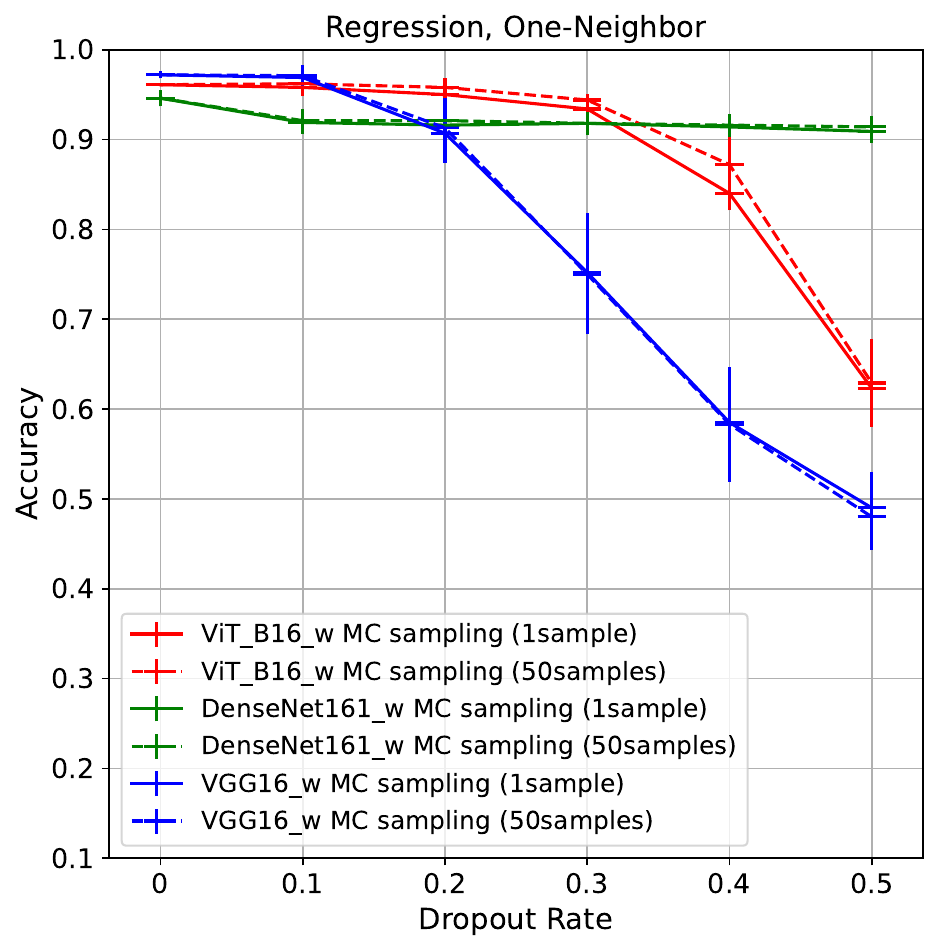}
    \label{fig:subfig_dropout_4}
  \end{subfigure}
  \caption{
    Exact class and one-neighbor class accuracy of each model at combined grade settings with varying dropout rates. 
  }
  \label{fig:appendix_dropout}
\end{figure}

Figure \ref{fig:appendix_dropout} depicts the exact class and one-neighbor class accuracy of the three deep learning models under the ablation study using different settings. The DenseNet and VGG model accuracy in classification settings showed less dependency on the dropout rate. This is thought to be due to the smaller number of Dropout layers compared to ViT. However, in regression, it can also be confirmed that VGG, like ViT, when the rate is higher, the accuracy is significantly reduced. The final values used in the validation experiments were enlisted in Table  \ref{tb:subsec_hyper_parameter}. 
\newpage
\captionsetup[sub]{font=tiny}
\section{Statistical tests}
\begin{figure}[h]
\centering
  \subcaptionbox{Combined, 50 samples}{%
    \includegraphics[width=0.5\textwidth]{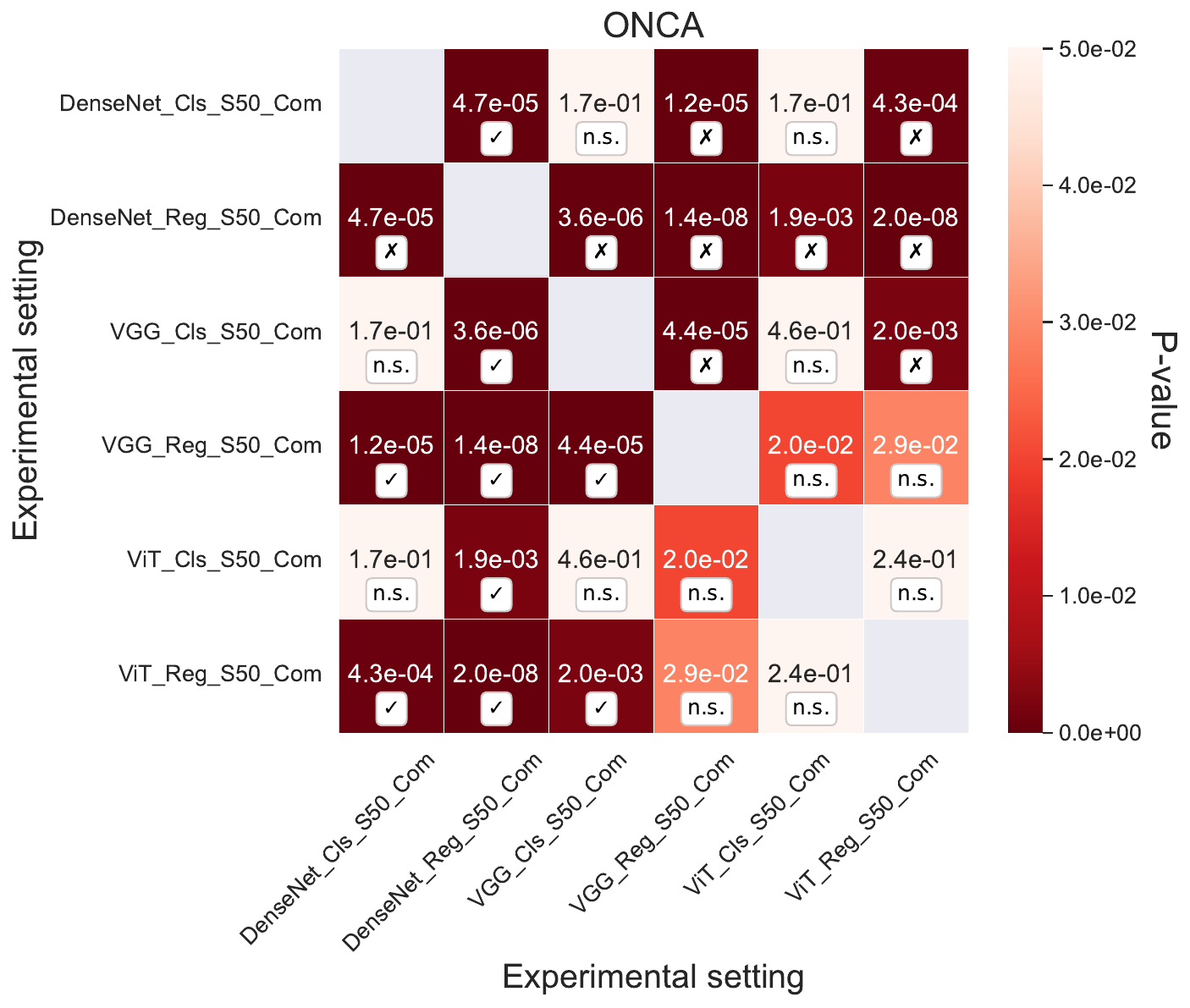}%
    \label{subfig_e}%
  }%
  \subcaptionbox{Separated, 50 samples}{%
    \includegraphics[width=0.5\textwidth]{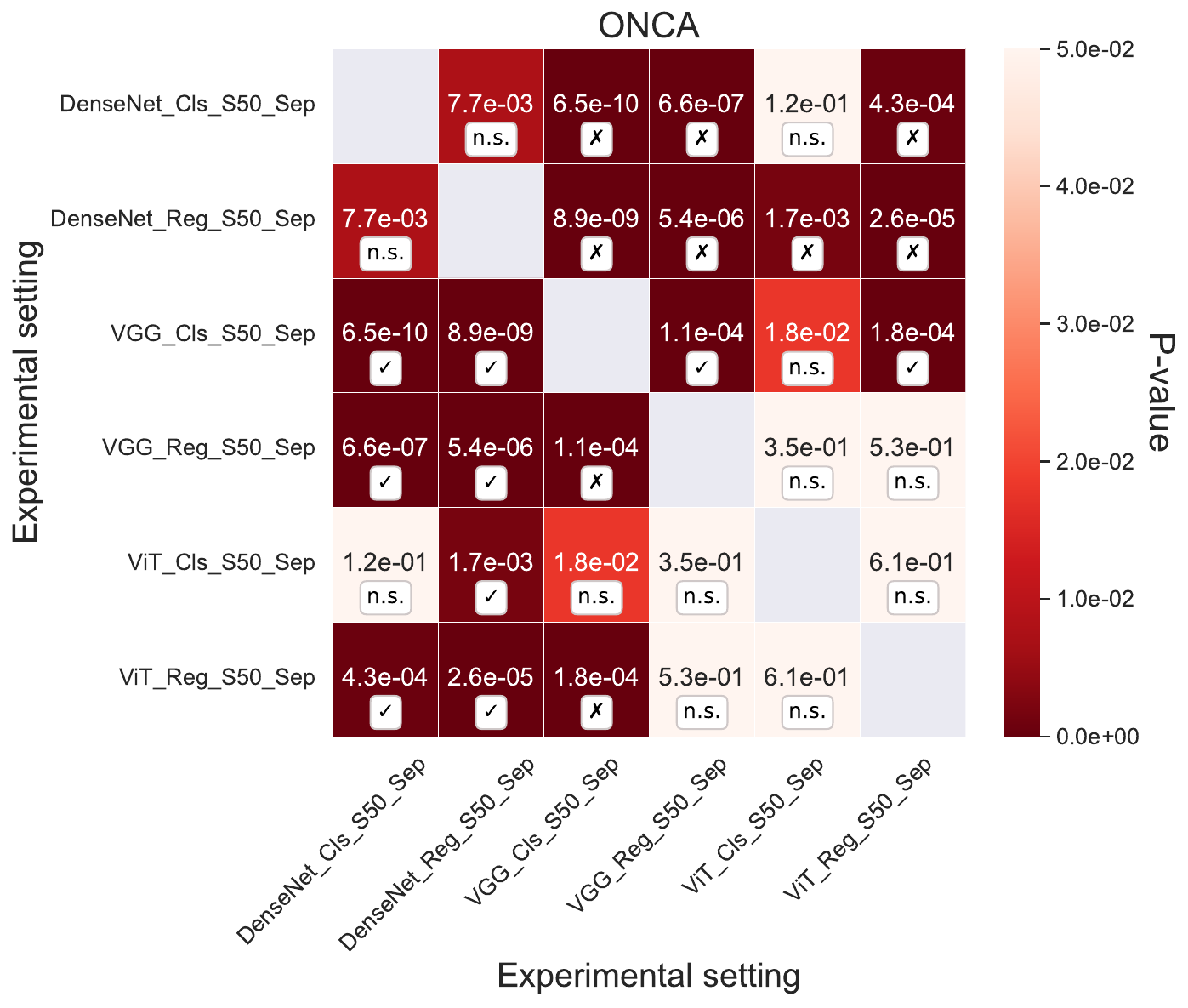}%
    \label{subfig_f}%
  }%
  \\
  \subcaptionbox{\hspace{0.5cm}Combined \newline \parbox{\linewidth}{\centering 1 vs 50 samples}}{%
    \includegraphics[width=0.25\textwidth]{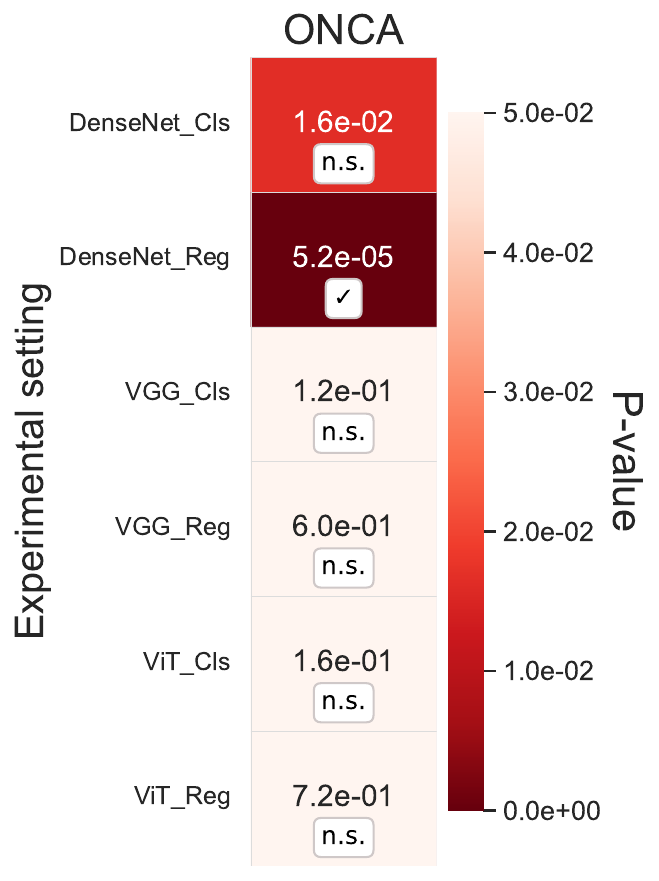}%
    \label{subfig_e}%
  }%
  \subcaptionbox{\hspace{0.5cm}Separated \newline \parbox{\linewidth}{\centering 1 vs 50 samples}}{%
    \includegraphics[width=0.25\textwidth]{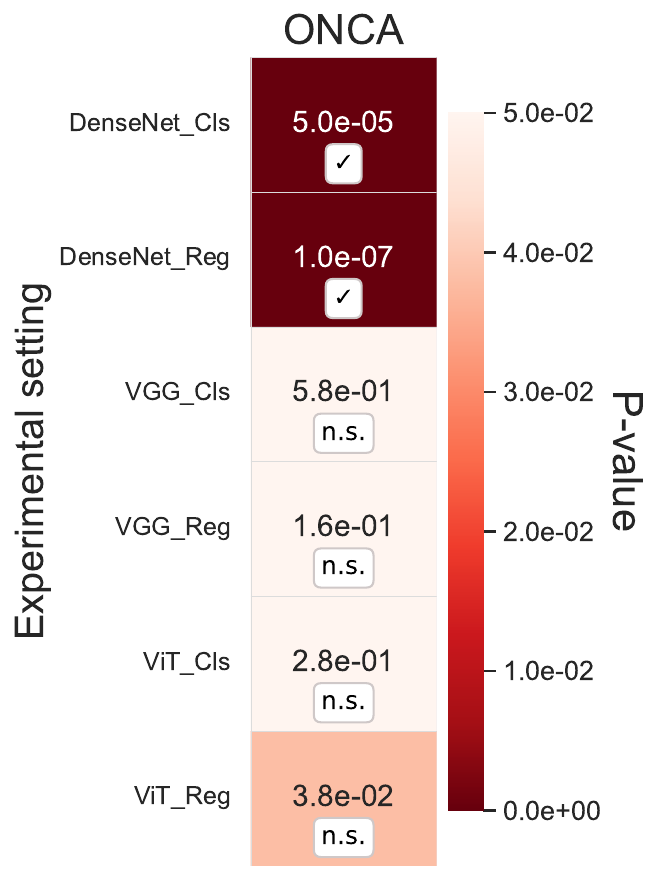}%
    \label{subfig_f}%
  }%
  \subcaptionbox{Combined vs Separated \newline \parbox{\linewidth}{\centering 50 samples}}{%
    \includegraphics[width=0.25\textwidth]{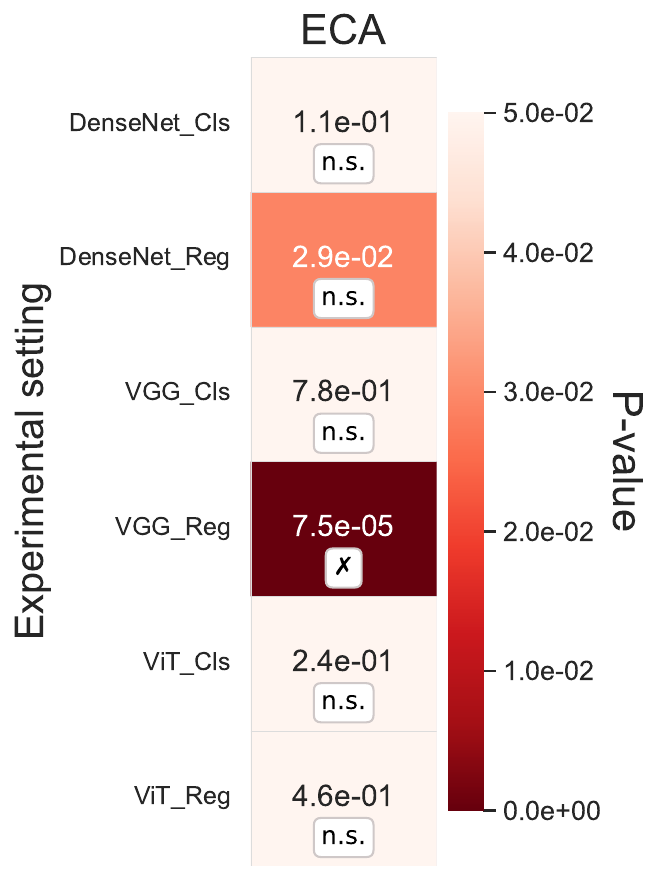}%
    \label{subfig_e}%
  }%
  \subcaptionbox{Combined vs Separated \newline \parbox{\linewidth}{\centering 50 samples}}{%
    \includegraphics[width=0.25\textwidth]{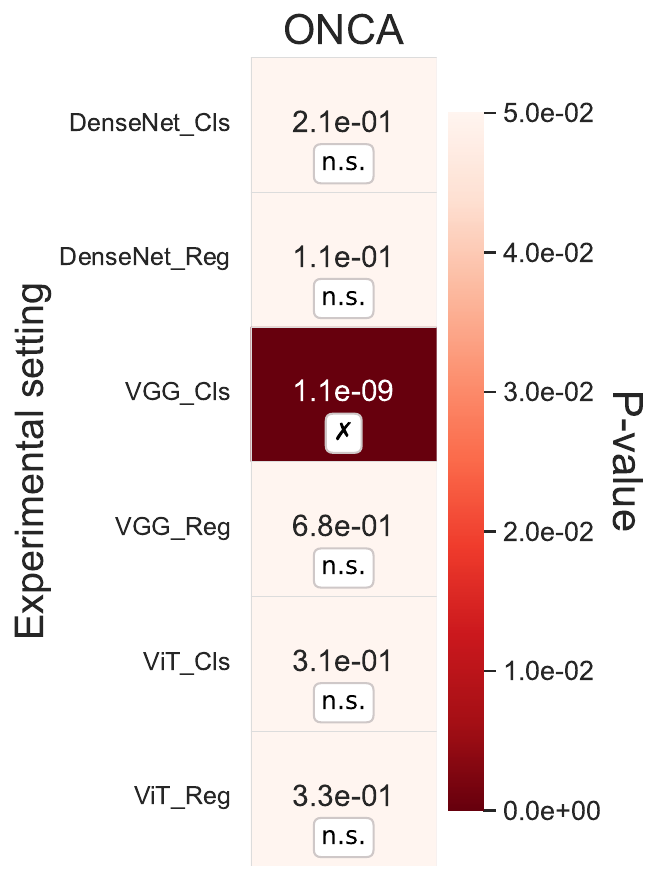}%
    \label{subfig_f}%
  }%

  \caption{P-values of the differences between the ECA of the three models and the prediction methods under the combined and separated labels, as well as the 1 and 50 samples settings. \textbf{\checkmark} in (a, b) indicates that the vertical experimental settings had higher accuracy than the horizontal setting; for (c, d, e, f), sample 1 setting had higher accuracy than the samples 50 setting with a statistically significant difference (Student's t-test with Bonferroni correction, corrected $\alpha$=3e-3 for (a, b), $\alpha$=1e-3 for (c, d, e, f). \textbf{\ding{55}} in (a, b) indicates that the vertical settings yielded lower accuracy; for (c, d, e, f), sample 1 yielded lower accuracy with a statistically significant difference, and \textbf{n.s.} when no significant difference was observed. Reg: regression, Cls: classification, S1: 1 samples (w/o dropout), S50: 50 samples (w/ dropout), Com: combined, Sep: separated.}
  \label{fig:appendix_pvalue}
\end{figure}

Figure \ref{fig:appendix_pvalue} depicts the P-values of the statistical tests (Student's t-test with Bonferroni correction) of the comparisons between the models under different settings.
\end{appendices}
\end{document}